\documentclass[twocolumn,english,aps,reprint, superscriptaddress,showpacs,showkeys]{revtex4-2}
\usepackage[colorlinks,citecolor=blue,urlcolor=blue,linkcolor = blue]{hyperref}
\usepackage{soul}

\usepackage[mathscr]{euscript}

\usepackage{hyperref}

\usepackage{graphicx}
\usepackage{cancel}
\usepackage{enumerate}
\usepackage[colorlinks,linkcolor=blue,citecolor=blue,urlcolor=blue]{hyperref}
\usepackage{textcomp}
\usepackage{amsmath}
\usepackage{amssymb}
\usepackage{pgf}
\usepackage{soul}
\usepackage{multirow}
\usepackage{lipsum}
\usepackage[normalem]{ulem}

\usepackage{comment} 

\usepackage[english]{babel}
\usepackage{bm}
\usepackage{tikz}
\usetikzlibrary{shapes.geometric}
\usetikzlibrary{decorations.markings}
\usetikzlibrary{decorations.pathmorphing,patterns}
\usetikzlibrary{decorations}
\usetikzlibrary{calc}
\usetikzlibrary{intersections}
\usetikzlibrary{arrows}
\usetikzlibrary{matrix}
\usetikzlibrary{positioning}

\newcommand{\balpha}{\boldsymbol{\alpha}}
\newcommand{\bbeta}{\boldsymbol{\beta}}
\newcommand{\bgamma}{\boldsymbol{\gamma}}

\newcommand{\nep}{{\mathrm{e}}}

\def\bra#1{\langle#1 |}
\def\ket#1{| #1\rangle}

\newcommand{\rmT}{\mathrm{T}}
\newcommand{\Fidelity}{\mathcal{F}}
\newcommand{\gs}{\mathrm{\scriptscriptstyle gs}}
\newcommand{\tot}{\mathrm{\scriptscriptstyle tot}}
\newcommand{\qopt}{\star}
\newcommand{\even}{\mathrm{\scriptscriptstyle even}}
\newcommand{\odd}{\mathrm{\scriptscriptstyle odd}}

\newcommand{\DeltaY}{\Delta_{\scriptscriptstyle \mathrm{Y}}}
\newcommand{\DeltaZ}{\Delta_{\scriptscriptstyle \mathrm{Z}}}

\newcommand{\XYZ}{\mathrm{\scriptscriptstyle XYZ}}

\newcommand{\target}{\mathrm{\scriptscriptstyle targ}}

\newcommand{\Spin}{\widehat{\mathrm S}}
\newcommand{\Parity}{\widehat{\mathrm P}}
\newcommand{\Inversion}{\widehat{\mathrm I}}
\newcommand{\Translation}{\widehat{\mathrm T}}
\newcommand{\PauliSigma}{\hat{\sigma}}

\newcommand{\Ho}{\widehat{H}}

\newcommand{\Uni}{\widehat{U}}
\newcommand{\Ham}{\widehat{H}}

\newcommand{\Ptrot}{\mathrm{P}}

\newcommand{\Tprod}[1]{\prod^{\leftarrow #1}}

\newcommand{\sym}{\widehat{S}}

\graphicspath{{./}{./Fig/}}

\begin{document}

\title{Avoiding barren plateaus via transferability of smooth solutions in Hamiltonian Variational Ansatz}

\author{Antonio A. Mele}
\affiliation{Dahlem Center for Complex Quantum Systems, Freie Universität Berlin, 14195 Berlin, Germany}
\affiliation{SISSA, Via Bonomea 265, I-34136 Trieste, Italy}
\author{Glen B. Mbeng}
\affiliation{Universit\"at Innsbruck, Technikerstra{\ss}e 21 a, A-6020 Innsbruck, Austria}
\author{Giuseppe E. Santoro}
\affiliation{SISSA, Via Bonomea 265, I-34136 Trieste, Italy}
\affiliation{International Centre for Theoretical Physics (ICTP), P.O.Box 586, I-34014 Trieste, Italy}
\affiliation{CNR-IOM, Consiglio Nazionale delle Ricerche - Istituto Officina dei Materiali, c/o SISSA Via Bonomea 265, 34136 Trieste, Italy}
\author{Mario Collura}
\affiliation{SISSA, Via Bonomea 265, I-34136 Trieste, Italy}
\affiliation{INFN Sezione di Trieste, via Bonomea 265, 34136 Trieste, Italy}
\author{Pietro Torta}
\affiliation{SISSA, Via Bonomea 265, I-34136 Trieste, Italy}

\begin{abstract}
A large ongoing research effort focuses on
Variational Quantum Algorithms (VQAs), representing leading candidates to achieve computational speed-ups on current quantum devices.
The scalability of VQAs to a large number of qubits, beyond the simulation capabilities of classical computers, is still debated. 
Two major hurdles are the proliferation of low-quality variational local minima, and the exponential vanishing of gradients in the cost function landscape, a phenomenon referred to as barren plateaus. 
In this work, we show that by employing iterative search schemes one can effectively prepare the ground state of paradigmatic quantum many-body models, also circumventing the barren plateau phenomenon. 
This is accomplished by leveraging the transferability to larger system sizes of a class of iterative solutions, displaying an intrinsic smoothness of the variational parameters, a result that does not extend to other solutions found via random-start local optimization.
Our scheme could be directly tested on near-term quantum devices, running a refinement optimization in a favorable local landscape with non-vanishing gradients.
\end{abstract}

\maketitle

%
%
{\em Introduction.---}
Variational Quantum Algorithms (VQAs) \cite{Cerezo_NEW,BhartiREV} are among the main candidates for near-term practical applications of Noisy Intermediate-Scale Quantum (NISQ) devices~\cite{Preskill_2018}.
VQAs are quantum-classical \emph{hybrid} optimization schemes that have been successfully applied to quantum ground state preparation~\cite{Ho_2019EfficientNonTrivial,AvoidingXXZtfimWierichs_2020, LGT_QAOA_2021} and classical optimization tasks~\cite{Farhi_arXiv2014}, ranging from the solution of linear systems of equations~\cite{BravoNew} to quantum information~\cite{VQAsQINFO}.
In the standard VQA setting, one aims at minimizing the average energy of a problem Hamiltonian $\Ho_{\target}$ with respect to a variational state $\ket{\psi\left(\bgamma\right)}$ prepared by a parameterized quantum circuit. This is accomplished by a feedback loop between a classical and a quantum machine: the quantum device is used to repeatedly prepare the ansatz state for a set of gate parameters $\bgamma$ and to estimate the cost function $E_{\textrm{var}}(\bgamma)=\langle \psi(\bgamma)|\Ho_{\target}| \psi(\bgamma)\rangle$, while the optimization of the parameters is performed classically.

The optimization of the cost function $E_{\textrm{var}}(\bgamma)$ is known to be a difficult task~\cite{Bittel_2021}: only a careful choice of the ansatz is usually \emph{expressive} enough to approximately find the ground state of $\Ho_{\target}$ and, at the same time, \emph{trainable} enough for the optimization to succeed. In particular, the landscape of the cost function may not be easy to inspect for two reasons: the proliferation of low-quality local minima traps~\cite{VQA_traps_NEW}, and the exponential flattening of the landscape by increasing the number of qubits, a phenomenon dubbed \emph{barren plateaus}~\cite{McClean_2018NEW}, which can severely hinder the scalability of the VQA scheme beyond small system sizes amenable to classical simulations. 
%
%
Barren plateaus are linked to highly-expressive parameterized quantum circuits~\cite{McClean_2018NEW, BPexpress_holmes2021connecting,EntInducedBP}, but they arise also in the context of less-expressive symmetry-preserving \cite{LaroccaDiagnBP,Wiersema_2020} or equivariant \cite{MyFirstPaper_SymmQML} ansatzes.
A few recent studies have proposed different approaches to limit or avoid barren plateaus, by employing pre-training techniques~\cite{MPS_QAOA_BP}, 
layerwise learning for classification tasks~\cite{Skolik_2021}, identity-block initialization~\cite{Grant_2019},
or classical shadows~\cite{BPshadows}. 

Among the effective strategies to avoid low-quality local minima traps, we mention approaches~\cite{SackAnnealingInitQAOA, QAOA_perceptron_NEW, LGT_QAOA_2021} inspired by standard Adiabatic Quantum Computation (AQC) \cite{AQC_Albash_2018,AQCfarhiNEW}, and iterative schemes~\cite{Zhou_PRX2020,Mbeng_arXiv2019}, optimizing only a subset of gate parameters at each iteration and using this result as a warm-start guess for the next iterative step.
These techniques proved particularly efficient for a class of VQAs
inspired by AQC, commonly named Hamiltonian Variational Ansatz (HVA)~\cite{Farhi_arXiv2014, VQAweckerPhysRevA.92.042303,Ho_2019EfficientNonTrivial,Ho_2019bUltrafast,MatteoWauthersPhysRevA.102.062404,Kokail_2019,Wiersema_2020,AvoidingXXZtfimWierichs_2020,MatosPRXQuantum.2.010309,SYMMBREAKINGqaoapark2021efficient, Ho_2019EfficientNonTrivial,CarleoVQEheis}, with an ansatz state of the form:
\begin{align} \label{eq:QAOA_like_ansatz}
\ket{\psi\left(\bgamma\right)}=\prod_{m=1}^{\Ptrot}\nep^{-i\gamma_{m,M}\Ham_{M}}\cdots\nep^{-i\gamma_{m,1}\Ham_{1}  }\ket{\psi_0} \;,
\end{align}
with $m=1\cdots\Ptrot$ labeling successive circuit layers, each in turn composed by $j=1\cdots M$ alternating unitaries generated by Hamiltonian operators $\Ham_{j}$. The target Hamiltonian $\Ho_{\target}$ can be linearly decomposed in terms of the generators, and $\ket{\psi_0}$
is a simple initial state.
%
This ansatz state can be regarded as a generalization of the Quantum Approximate Optimization Algorithm (QAOA)~\cite{Farhi_arXiv2014}, originally devised for classical combinatorial optimization problems.
Remarkably, by means of appropriate iterative schemes for constructing the layer parameters $\gamma_{m,j}$, it is often possible to efficiently single-out optimal or nearly-optimal variational parameters that are \emph{smooth} functions~\cite{Mbeng_PhDThesis2019New,Pagano_2020,Zhou_2020, farhi2020quantum,Wurtz_2022,Crooks_arXiv2018,QAOApatternbrady2021behavior} of the layer index $m$.

In this paper, we draw a new connection between smooth optimal solutions --- obtained by means of iterative methods --- and barren plateaus, developing a novel efficient scheme to circumvent this issue. 
%
Our procedure leverages the \emph{transferability} of an optimal \emph{smooth} solution, obtained for small system size, to solve the same task with a larger number of qubits, where a direct optimization would fail due to barren plateaus. 
In a nutshell, the transferred smooth solution serves as an excellent warm-start with low variational energy for the large system, and a subsequent refinement optimization is observed to be free of the barren plateau issue. 
Remarkably, even though other 
(non-smooth) solutions for the small system can be obtained by standard random-start local optimization, they do not provide any useful warm-start for larger systems and, crucially, a refinement optimization still suffers from barren plateaus in their neighborhood.

For definiteness, we focus on the ground state preparation of the Heisenberg XYZ model \cite{book:Baxter} and of the antiferromagnetic Longitudinal-Transverse-Field Ising Model (LTFIM) \cite{LTFIM2003}, two ubiquitous models in quantum physics with rich phase diagrams, whose ground state preparation with VQAs is affected by barren plateaus~\cite{LaroccaDiagnBP,MyFirstPaper_SymmQML}.
We select ansatz states in the form of Eq.~\eqref{eq:QAOA_like_ansatz}, by choosing the generators $\Ham_{j}$ in such a way to implement model symmetries into the variational wavefunctions. 
This leads to a restriction of the Hilbert space to the ground state symmetry sector, boosting trainability, and a reduction in the number of independent Pauli correlators needed to compute the cost function.


%
{\em Models and methods.---} The first class of models we consider is the spin-1/2 XYZ~\cite{book:Baxter,Baxter_AnnPhys1972} Hamiltonian:
\begin{equation}
\label{eq:XYZ_Ham}
\Ham_{\XYZ} = \sum_{j=1}^{N} 
\Big( \PauliSigma_{j}^{x} \PauliSigma_{j+1}^{x} + 
\DeltaY \PauliSigma_{j}^{y} \PauliSigma_{j+1}^{y} +
\DeltaZ \PauliSigma_{j}^{z} \PauliSigma_{j+1}^{z} \Big) \;.
\end{equation}
We restrict our considerations to the antiferromagnetic case $\DeltaY, \DeltaZ>0$. In this quadrant, the system is gapped, except at three critical half-lines/segments: $\DeltaY\leq 1,\DeltaZ=1$; $\DeltaY=1,\DeltaZ\leq 1$; $\DeltaY=\DeltaZ,\DeltaZ \geq 1$ ~\cite{denNijs_PRB1981,Ercolessi_PRB2011}. The Hamiltonian~\eqref{eq:XYZ_Ham} is integrable in the whole $(\DeltaY, \DeltaZ)$ plane. In particular, $\DeltaY=1$ corresponds to the XXZ model, while $\DeltaY=\DeltaZ=1$ corresponds to the spin-isotropic Heisenberg model.
The second Hamiltonian we examine is the antiferromagnetic LTFIM~\cite{Ovchinnikov_PRB2003,Sen_PRE2001}:
\begin{equation} \label{eq:LTFIM_Ham}
    \widehat{H}_{\scriptscriptstyle\mathrm{LTFIM}}=
    \sum_{j=1}^{N}\hat{\sigma}_{j}^{z} \hat{\sigma}_{j+1}^{z} - g_x\sum_{j=1}^{N}\hat{\sigma}_{j}^{x} - g_z\sum_{j=1}^{N}\hat{\sigma}_{j}^{z }\;.
\end{equation}
We restrict our analysis to positive local fields $g_x, g_z>0$.
The system is gapped in the whole positive quadrant, except for a line connecting the two points $(g_x=1, g_z=0)$ and $(g_x=0, g_z=2)$,
obtained numerically in Ref.~\cite{Ovchinnikov_PRB2003}. 
While for $g_z=0$ the model is integrable by a Jordan-Wigner transformation to free fermions~\cite{Jordan_ZeitPhys1928,Leib_AnnPhys1961,MbengDigANN,AvoidingXXZtfimWierichs_2020,WangFermQAOA}, integrability is generically lost for $g_z\neq 0$.  
In Appendix~\ref{app:TFIM}, we specifically address the integrable Transverse Field Ising Model (TFIM) line $(g_z=0)$.
%
For both the XYZ model and the LTFIM, we examine even values of $N$ and we assume periodic boundary conditions.

%
Our ansatz states are in the general form of Eq.~\eqref{eq:QAOA_like_ansatz} with $M=2$ generating Hamiltonians only, defined to encode some symmetries of the model.
To illustrate this idea for the XYZ case, let us split $\Ham_{\XYZ}$ into two mutually non-commuting parts that refer to the {\em even} 
$(2j-1,2j)$ and to the {\em odd} bonds $(2j,2j+1)$, 
$\Ham_{\XYZ}= \Ham_{\even} + \Ham_{\odd}$, with
\begin{equation} \label{eq:even_Ham}
\Ham_{\even} = \sum_{j=1}^{N/2} 
\Big(\PauliSigma_{2j-1}^{x} \PauliSigma_{2j}^{x} + \DeltaY 
\PauliSigma_{2j-1}^{y} \PauliSigma_{2j}^{y} + \DeltaZ
\PauliSigma_{2j-1}^{z} \PauliSigma_{2j}^{z}\Big)
\end{equation}
and similarly for $\Ham_{\odd}$.
Next, in the spirit of AQC~\cite{AQC_Albash_2018}, imagine an {\em interpolating Hamiltonian} connecting $\Ham_{\even}$ to the full 
$\Ham_{\XYZ}$:
\begin{equation}
\label{eq:adiab_Ham}
\Ham(s) = s \Ham_{\XYZ} + (1-s) \Ham_{\even} = \Ham_{\even} + s \Ham_{\odd} \;,
\end{equation}
with $s\in [0,1]$.
For $s=0$, the ground state of $\Ham(0)=\Ham_{\even}$ is a valence-bond state of singlets on the even bonds
\begin{equation} \label{eq:init_state_XYZ}
\ket{\psi_0} = \prod_{j=1}^{N/2} \frac{1}{\sqrt{2}} \Big( \ket{\!\uparrow\downarrow}-\ket{\!\downarrow\uparrow} \Big)_{2j-1,2j} \;,
\end{equation}
which is taken as initial state.
This suggests, in close analogy with QAOA, the following ansatz for the XYZ ground-state wave-function:
\begin{equation} \label{eq:ansatz_state}
\ket{\psi(\bbeta,\balpha)_{\Ptrot}} = \Uni_{\Ptrot} \cdots \Uni_2 \,  \Uni_1 \ket{\psi_0} \;.
\end{equation}
%
%
Here, $(\bbeta,\balpha)_{\Ptrot} = (\beta_1\cdots\beta_{\Ptrot}\,, \alpha_1\cdots\alpha_{\Ptrot})$ are $2\Ptrot$ variational parameters, and the unitary operators $\Uni_m=\Uni(\beta_{m},\alpha_{m})$, for $m=1\cdots \Ptrot$, evolve the state according to $\Ham_{\even}$ and $\Ham_{\odd}$, in an alternating fashion:
\begin{equation} \label{eq:ansatz_evolution_XYZ}
    \Uni_m = \Uni(\beta_m,\alpha_m) = \nep^{-i\beta_m \Ham_{\even}} \nep^{-i\alpha_m \Ham_{\odd}} \;.
\end{equation}

As usual in the VQA framework, the goal is to minimize the variational energy
\begin{equation} \label{eq:cost_function}
    E_N \left(\bbeta,\balpha \right)_\Ptrot = \langle \psi(\bbeta,\balpha)_{\Ptrot} | \Ho_{\target} | \psi(\bbeta,\balpha)_{\Ptrot} \rangle \;,
\end{equation}
with $\Ho_{\target}=\Ham_{\XYZ}$.
The connection with AQC is restored in the $\Ptrot\to\infty$ limit, by setting specific values for $(\bbeta,\balpha)_\Ptrot$, as prescribed by a Trotter split-up of the continuous-time AQC dynamics~\cite{Mbeng_arXiv2019}.

As detailed in Appendix~\ref{app:models_and_ansatzs}, the ansatz state lies in the same symmetry subsector of the XYZ ground states for the following symmetries: translations by {\em two} lattice spacings $\Translation^2$ (which maps $j\to j+2$), lattice inversion $\Inversion$ (which maps $j\leftrightarrow N-j+1$) and parity $\Parity_b=\prod_j \PauliSigma^b_j$. Additionally, for the su(2)-invariant Heisenberg model, this holds true for the total spin $\Spin_{\tot}^b$ ($b=x,y,z$) and
$\Spin_{\tot}^2$, while for the u(1)-invariant XXZ model only for $\Spin_{\tot}^z$.
As a result, the cost function in Eq.~\eqref{eq:cost_function} requires the evaluation of only six independent two-points correlators, which may be further reduced to four (two) by exploiting rotational symmetries in the XXZ (XXX) case. 

The ansatz 
for the ground state preparation of the LTFIM reads as in Eq.~\eqref{eq:ansatz_state}, with a single layer unitary given by:
\begin{equation} \label{eq:ansatz_evolution_LTFIM}
    \Uni_m = \nep^{i\beta_m \Ham_{\scriptscriptstyle \mathrm{X}} } \nep^{-i\alpha_m \left(\Ham_{\scriptscriptstyle\mathrm{ZZ}} - g_z \Ham_{\scriptscriptstyle\mathrm{Z}} \right)}  \;,
\end{equation}
where we defined $\Ham_{\scriptscriptstyle\mathrm{ZZ}}$, $\Ham_{\scriptscriptstyle\mathrm{Z}}$, and $\Ham_{\scriptscriptstyle \mathrm{X}}$ simply as the sum of nearest-neighbors interactions, Pauli-z and Pauli-x operators, respectively.
In this setting, the initial state is simply the fully-polarized state along-x $\ket{\psi_0} = \ket{+}^{\otimes N}$,
once again bearing a direct connection with AQC state preparation for $\Ptrot\to\infty$.
The goal is to minimize the variational energy as in Eq.~\eqref{eq:cost_function}, now with $\Ho_{\target}=\widehat{H}_{\scriptscriptstyle\mathrm{LTFIM}}$. 
Also in this case (Appendix~\ref{app:models_and_ansatzs}), the variational ansatz is restricted to the correct symmetry subsector of the target ground state, for both single-site translation $\widehat{\operatorname{T}}$ (full translational invariance) and lattice inversion $\Inversion$.
%

A natural interpretation in terms of light-cone spreading of quantum correlations emerges for both our ansatz wavefunctions (see Appendix~\ref{app_light-cone}). 
As a main consequence, the whole cost function landscape, once rescaled by the system size $N$,
becomes independent of $N$ itself for $N>\tilde{N}_\Ptrot$, where $\tilde{N}_\Ptrot=4\Ptrot+2$ and $\tilde{N}_\Ptrot=2\Ptrot+1$ for the XYZ and the LTFIM, respectively. 
%

{\em Results.---} 
In this work, we adopt an iterative interpolation scheme (INTERP)~\cite{Zhou_PRX2020,Mbeng_arXiv2019} which was originally formulated for standard QAOA applied to classical optimization problems.
Here, we apply this heuristic to more general HVA wavefunctions as in Eq.~\eqref{eq:QAOA_like_ansatz}, with the goal of quantum many-body ground state preparation.
Essentially, the idea is to perform a sequence of local optimizations for increasing values of $\Ptrot$, each of them starting from an educated guess that is iteratively updated, by interpolating on the optimal parameters found at the previous step.
Additional details on this algorithm are reported in Appendix~\ref{app:additional_numerical_res}, where we also provide numerical evidence that both XYZ and LTFIM ground states can be efficiently prepared 
across their phase diagrams, reaching high fidelity values.

\begin{figure}[t]
\centering
\includegraphics[width=0.47\textwidth]{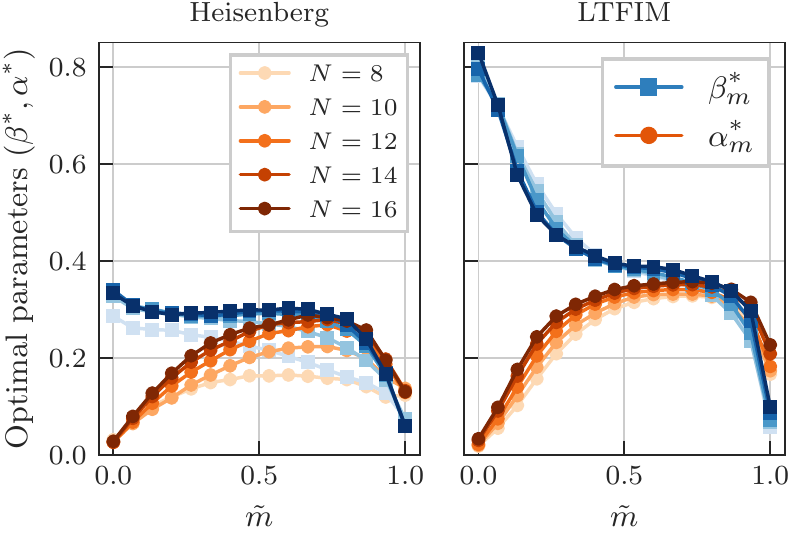}
\caption{Smooth optimal parameters $(\bbeta^*,\balpha^*)\rvert_{\Ptrot,N} = (\beta^*_1\cdots\beta^*_{\Ptrot}\,, \alpha^*_1\cdots\alpha^*_{\Ptrot})$ obtained with INTERP (see main text), plotted vs the rescaled index $\tilde{m}\equiv(m-1)/(\text{P}-1)$ in the $x$-axis range $[0,1]$.
Results are shown for the Heisenberg model ($\DeltaY=1$, $\DeltaZ=1$) (left) and the LTFIM ($g_x=1$, $g_z=1$) (right) for
$\Ptrot=16$, and they are qualitatively similar for different sizes $N$.
These solutions are stable by further increasing the number of layers $\Ptrot$.
Similar smooth solutions can be found for different points of the phase diagram.}
\label{fig:smooth_curves}
\end{figure}

\begin{figure}[t]
\centering
\includegraphics[width=0.475\textwidth]{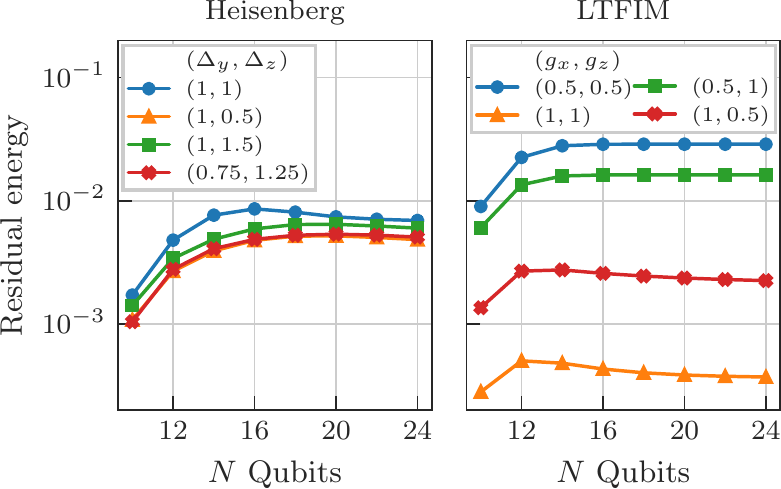}
\caption{Residual energies (Eq.~\eqref{eq:residual_energy}) for different system sizes using parameters from a small-size ``guess-system" ($N_G=8$) computed for different flavours of the two models.
We show the transferability of smooth optimal solutions $(\bbeta^*, \balpha^*)\rvert_{\Ptrot, N_G}$ with $\Ptrot=10$ for XYZ models (left) and the LTFIM (right).}
\label{fig:transferibility}
\end{figure}

Usually, by adopting such iterative methods, one finds optimal angles that are smooth functions of the layer index $m=1\cdots\Ptrot$. For this reason, we dub them \emph{smooth} solutions. 
This is consistently observed in all phases of our models, as shown in Fig.~\ref{fig:smooth_curves} at the critical point of XXZ (Heisenberg model) and close to the critical line of the LTFIM~\cite{Ovchinnikov_PRB2003}. On top of that, we note that these smooth optimal curves are qualitatively similar for different system sizes.
Inspired by this observation, we verify numerically that smooth optimal solutions $(\bbeta^*, \balpha^*)\rvert_{\Ptrot,N_{G}}$ --- obtained by applying INTERP to a small-size system with dimension $N_G$ up to a certain value of $\Ptrot$ --- can be transferred to solve the same task for a larger number of qubits, thus providing an effective educated guess.
In the following, we will always indicate with $N_{G}$ the ``guess" size used to obtain the optimal smooth solution, which will be eventually transferred to a larger system with $N>N_{G}$ lattice sites. Unless otherwise stated, we set $N_{G}=8$.
In order to estimate the effectiveness of our transferability protocol, we define the \emph{residual energy} as
%
\begin{equation}
    \label{eq:residual_energy}
    \varepsilon_{N}(\bbeta^*, \balpha^*)\rvert_{\Ptrot,N_G}
    =  \frac{E_N (\bbeta^*, \balpha^*)\rvert_{\Ptrot,N_G} -E^\text{min}_N}{E^\text{max}_N-E^\text{min}_N} \;,
\end{equation}
where $E_N (\bbeta^*, \balpha^*)\rvert_{\Ptrot,N_G}$ is the cost function in Eq.~\eqref{eq:cost_function} for a system of size $N$, evaluated at fixed angles $(\bbeta^*, \balpha^*)\rvert_{\Ptrot,N_{G}}$, 
while $E^\text{min}_N$ ($E^\text{max}_N$) is the ground-state (maximum) exact energy for such a size $N$.
In Fig.~\ref{fig:transferibility} we plot this quantity, for different points of the phase diagram of our models: strikingly, smooth optimal curves obtained for a small system 
provide an excellent educated guess for the ground-state preparation up to $N=24$ lattice sites.

A few comments are in order. The residual energy is usually a good proxy for the fidelity with the ground state. It may roughly evaluate at $\approx 0.5$ when computed at a random point in the energy landscape, while its values obtained via transferability are remarkably low. 
A more detailed study on the actual fidelity of transferred solutions with target ground states is carried out in Appendix~\ref{app:additional_numerical_res}.
Secondly, the transferability of this class of smooth solutions found via INTERP holds true for larger values of $\Ptrot$; in contrast, other equal-quality \emph{non-smooth} solutions for the small $N_G$-size system --- obtained by means of random-start local optimization --- do not provide any useful guess for the ground state preparation of the same model with a larger number of qubits. 
These results are reported and analyzed in Appendix~\ref{app:additional_numerical_res}.
Finally, we tested the existence of smooth curves, and their transferability to a larger number of qubits, also for the TFIM:
our results are confirmed up to much larger sizes, by leveraging a standard mapping to free fermions~\cite{Jordan_ZeitPhys1928,Leib_AnnPhys1961,MbengDigANN}, as reported in Appendix~\ref{app:TFIM}.

Despite the good educated guess provided by the transferability of smooth solutions, one may be tempted to refine the ground state approximation for the $N$-size model, e.g. by aiming at a target value of fidelity such as $99.9\%$. However, for such large sizes, both the XYZ models and the LTFIM are affected by barren plateaus~\cite{LaroccaDiagnBP,MyFirstPaper_SymmQML}. Therefore, any local optimization starting from a random point in the parameter space is doomed to fail on a realistic quantum device, due to vanishingly small gradients requiring an exponential scaling of resources~\cite{McClean_2018NEW,Cerezo_2021loc}.
Remarkably --- and this is the main novel result of our paper --- we find that transferred smooth optimal solutions stand out in this respect: in their neighborhood, the landscape does not suffer from small gradients, and a local optimization would succeed. 

\begin{figure}
\centering
\includegraphics[width=0.475\textwidth]{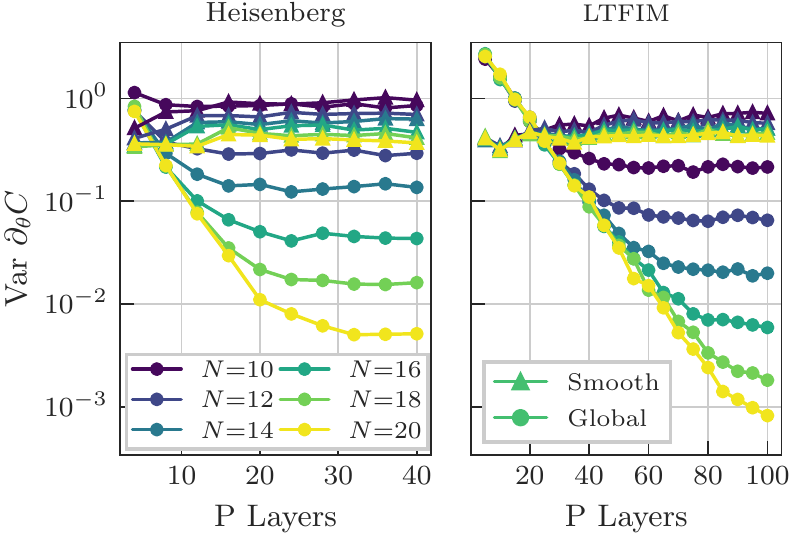}
\caption{Barren plateaus in the whole search space (data denoted as ``Global"), contrasted with a qualitatively different trend in the $\epsilon$-neighborhood of the transferred smooth solution $(\bbeta^*, \balpha^*)\rvert_{\Ptrot,N_G}$, obtained with INTERP for a small system size $N_G=8$ (data denoted as ``Smooth").
Here, we focus on a single partial derivative w.r.t. $\theta=\alpha_1$ (see Eqs.~\eqref{eq:ansatz_evolution_XYZ},~\eqref{eq:ansatz_evolution_LTFIM}) of the cost function in Eq.~\eqref{eq:cost_function}, rescaled as in Appendix~\ref{app:models_and_ansatzs} (see Eqs.~\eqref{eq:var_en_XYZ},~\eqref{eq:var_en_LTFIM}) and dubbed $C$.
We plot the sample variance of the partial derivative as a function of the number $\Ptrot$ of HVA layers in the circuit.  
We fix $\epsilon=0.05$ and a batch of $1000$ samples for each value of $\Ptrot$ and $N$.}
\label{fig:BP1}
\end{figure}

\begin{figure}[ht]
\centering
\includegraphics[width=0.475\textwidth]{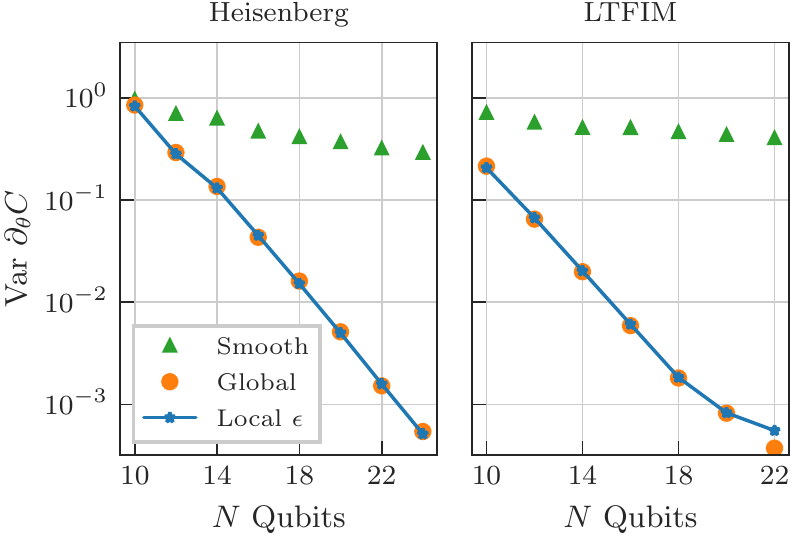}
\caption{Same quantity as in Fig.~\ref{fig:BP1}, here plotted vs the qubit number for a fixed circuit depth $\Ptrot=40$ ($\Ptrot=100$) for the Heisenberg model (LTFIM).
The labels ``Global" and ``Smooth" refer to the same data of Fig.~\ref{fig:BP1}.
We denote with ``Local $\epsilon$" a sampling performed in an $\epsilon$-region centered around a random point. Data are averages over $20$ random points (generated independently for each value of $N$) with $\epsilon=0.05$. We use a constant batch size of $1000$ samples, for each $N$ and sampling region.
A clear trend appears: the neighborhood of a random point (or that of a transferred non-smooth solution, see Appendix~\ref{app:additional_numerical_res}) exhibits the same exponential decay as the whole space, whereas this is not present in the neighborhood of the transferred smooth solution.}
\label{fig:BP2}
\end{figure}

Figures~\ref{fig:BP1} and~\ref{fig:BP2} illustrate this important point. For conciseness, we show data for ($\DeltaY=1$, $\DeltaZ=1$) and ($g_x=1$, $g_z=1$), but our results extend to other points of the phase diagrams. 
Specifically, in Fig.~\ref{fig:BP1}, we plot the variance of a representative gradient component of the cost function in Eq.~\eqref{eq:cost_function}, as customary in studies on barren plateaus~\cite{McClean_2018NEW,Wiersema_2020,BPexpress_holmes2021connecting}, which is sampled at random in the whole landscape. As expected, its exponential decay with the system size $N$ confirms the presence of barren plateaus. 
However, if we sample the same gradient component only in a neighborhood of radius $\epsilon$ of the transferred smooth solution, its magnitude does not show any appreciable exponential decay. This result is clearly observed for both classes of models under exam, and it is further evidenced in Fig.~\ref{fig:BP2}, showing data for a fixed value of $\Ptrot$: the exponential decrease of the gradients in the whole search space is equivalent to that in a neighborhood of radius $\epsilon$ of any given set of angles, with the exception of the smooth transferred curve $(\bbeta^*, \balpha^*)\rvert_{\Ptrot,N_G}$.
Once more, also this local-landscape property does not extend to the neighborhood of other transferred non-smooth solutions, which neither provide a useful educated guess for the large system nor solve the barren plateau issue for a local optimization. 
This is shown in Appendix~\ref{app:additional_numerical_res}, along with data supporting the effectiveness of a refinement optimization, performed classically in the neighborhood of the transferred smooth curve.

Incidentally, for each value of $N$, the sample variance in the whole search space saturates after a certain circuit depth $\Ptrot$, as argued in \cite{McClean_2018NEW,Cerezo_2021loc} and clearly shown in Fig.~\ref{fig:BP1}. This fact is usually linked to the ansatz parameterized quantum circuit approaching an approximate 2-design~\cite{Gross_2007,Brand_o_2016,HaferTdesign} on its symmetry subspace~\cite{LaroccaDiagnBP}.

Finally, let us remark that a direct application of the INTERP algorithm to a large $N$-size system might circumvent the barren plateaus issue for this class of problems: as shown in Appendix~\ref{app:additional_numerical_res}, the smooth optimal curve provides an effective warm start at each iteration of the algorithm, starting from low circuit depths up to deep circuits --- where a randomly-initialized local search would fail, due to barren plateaus.
However, it is manifestly more resource-efficient to apply INTERP to a small-size system $N=N_G$, and to leverage the transferability of smooth optimal solutions, as well as the absence of barren plateaus in their neighborhood. 
Indeed, our findings pave the way to an improved scheme to prepare the ground state of this class of many-body quantum systems with a large number of qubits:  
the smooth optimal curves can be found \emph{classically} for a small system, and then transferred to solve the same task for larger $N$, beyond the reach of classical simulations. The \emph{quantum} device would only be needed for a refinement optimization, in the absence of barren plateaus.

{\em Conclusions.---} 
We tackled many-body ground state preparation via problem-inspired VQAs, and provided extensive numerical evidence on the transferability of a class of optimal \textit{smooth} solutions --- obtained by means of iterative schemes for a small number of qubits --- to solve the same task for larger system sizes.
Remarkably, these solutions 
provide an excellent educated guess for the ground-state wave function, as opposed to other 
solutions that can be easily obtained for small systems without appropriate iterative schemes. 
These results are confirmed up to larger sizes for the TFIM.

Our procedure overcomes the well-known (and not yet fully addressed) difficulties related to the highly non-trivial structure of the variational energy landscape. On top of avoiding low-quality local minima traps daunting random-start local optimization~\cite{VQA_traps_NEW, Zhou_PRX2020}, we provided evidence of a novel and remarkable feature of this class of solutions:
the cost-function landscape is observed to become free of barren plateaus in their neighborhood, potentially allowing for further effective refinement optimizations with a quantum device on a classically-obtained smooth guess.

This work paves the way to a plethora of novel exciting research directions. Our new effective way of approaching ground state preparation for larger many-body systems may allow to deal with 2D lattice models, ranging from spin systems to Hubbard-like systems, with or without disorder.
%
%
On a theoretical side, it might be interesting to prove analytically the transferability and landscape properties of smooth solutions found via INTERP.
Previous numerical and analytical results on reusable optimal variational parameters exist, either among typical instances of a problem or across different system sizes~\cite{Akshay_2021,VerdonTransf,galda2021transferability,brandao2018fixed,Pagano_2020,Streif_2020}. 
These ``parameter concentration" results are usually limited to shallow circuits, while here we focus on smooth optimal solutions for large values of $\Ptrot$, providing a link between solution transferability and the local absence of barren plateaus.
A possible connection between this class of solutions and adiabaticity might be investigated~\cite{Mbeng_2019annealing,QAOApatternbrady2021behavior,Wurtz_2022}.
Finally, our scheme could be directly tested with near-term technology on real quantum devices, beyond the size limits of classical computation.

{\em Acknowledgments.---} We thank Johannes Jakob Meyer, Sumeet Khatri, Ryotaro Suzuki, Yihui Quek, Janek Denzler, Jakob S. Kottman and Jens Eisert for useful discussion.
The research was partly supported by EU Horizon 2020 under ERC-ULTRADISS, Grant Agreement No. 834402. AAM acknowledges support from BMBF (FermiQP and Hybrid).  GBM acknowledges support from Austrian Science Fund through the
SFB BeyondC Project No. F7108-N38.
GES acknowledges that his research has been conducted within the framework of the Trieste Institute for Theoretical Quantum Technologies (TQT). 

\bibliography{BiblioQAOA, BiblioLRB, BiblioQA_ter, BiblioQIC, BiblioQIsing, BiblioQSL, Bib_PT, BIBLIO_ThesisAntMele, Biblio_Ant_New}

\appendix

\counterwithin{figure}{section}
\counterwithin{table}{section}
\renewcommand\thefigure{\thesection\arabic{figure}}
\renewcommand\thetable{\thesection\arabic{table}}

\section{Symmetries and ansatz wavefunctions}
\label{app:models_and_ansatzs}

\subsection{Symmetries encoded in the HVA}

An important aspect of the story concerns the symmetries of the ground state we need to construct and those of our ansatz wavefunction.
In this section, we describe in greater detail the relevant symmetries of the XYZ models and the LTFIM. 
%
We will show that the choice of ansatz wavefunctions as in Eqs.~\eqref{eq:ansatz_state},~\eqref{eq:ansatz_evolution_XYZ},~\eqref{eq:ansatz_evolution_LTFIM} restricts the search space from the whole Hilbert space to a specific symmetry sub-sector, which is precisely the one where the target ground state belongs.

Let us start with general considerations valid for problem-inspired HVA in Eq.~\eqref{eq:QAOA_like_ansatz}.
Suppose we identify a set of symmetries of the target Hamiltonian $\Ho_{\target}$ --- whose ground state we aim to prepare --- and let us focus in particular on a specific symmetry (unitary) operator $\sym$. 

Using the same notation as in the main text,
a smart strategy can be to select the generators $\widehat{H}_1,\dots,\widehat{H}_M$ such that they \emph{all} commute with $\sym$.
In fact, if we select as initial state an easy-to-prepare symmetry eigenstate $\sym\ket{\psi_0}=e^{i\phi}\ket{\psi_0}$, it immediately follows that the HVA is confined to the same symmetry subsector, since:
\begin{equation} \label{eq:symmetry_sector}
         \sym\ket{\psi\left(\bgamma\right)}=e^{i\phi}\ket{\psi\left(\bgamma\right)}\;,
\end{equation}
for any choice of the variational parameters.
The rationale behind this procedure is simple: we should select the correct symmetry subsector, where the target ground state belongs, by properly choosing $\ket{\psi_0}$: this sector is then preserved by applying only symmetry-commuting unitaries.

For the sake of clarity, let us now restrict our discussion to the XYZ model; the LTFIM requires only minor changes summarized at the end of this section.
As a preliminary observation, we remark that the various bond terms appearing in $\Ham_{\even}$ (or equivalently, in $\Ham_{\odd}$) form a set of mutually commuting operators, hence the corresponding unitaries factorize; even more, since $\left[\PauliSigma_{j}^{b} \PauliSigma_{j+1}^{b},\PauliSigma_{j}^{b'} \PauliSigma_{j+1}^{b'}\right]=0$ for $b,b'=x,y,z$, the various unitaries also factorize in the $xx$, $yy$ and $zz$ terms. 
This leads to a standard parameterized quantum circuit, which can be further decomposed into a basis set of native gates (e.g. CNOT and single-qubit rotations)~\cite{Nielsen_Chuang:book}.
Let us simplify the notation for this ansatz state, evaluated in a generic point of the search space, by setting $\ket{\psi_{\Ptrot}}\equiv \ket{\psi(\bbeta,\balpha)_{\Ptrot}}$. 

The initial state $\ket{\psi_0}$ in Eq.~\eqref{eq:init_state_XYZ} has obvious symmetries with respect to translations by {\em two} lattice spacings $\Translation^2$ (which sends $j\to j+2$), lattice inversion $\Inversion$ (which maps $j\leftrightarrow N-j+1$), parity $\Parity_b=\prod_j \PauliSigma^b_j$  and total spin $\Spin_{\tot}^b$, with $b=x,y,z$, as well as 
$\Spin_{\tot}^2$. 
Clearly, $\Spin_{\tot}^2\ket{\psi_0}=0$, and $\Spin_{\tot}^b\ket{\psi_0}=0$.
The singlets, however, are {\em odd} under exchange of the two spins and also under application of $\Parity_b$. 
Hence, while $\Translation^2\ket{\psi_0} = \ket{\psi_0}$, we have that $\Parity_b \ket{\psi_0} = (-1)^{\frac{N}{2}} \ket{\psi_0}$ and $\Inversion \ket{\psi_0} = (-1)^{\frac{N}{2}} \ket{\psi_0}$.

Concerning the symmetries of the ansatz state $\ket{\psi_{\Ptrot}}$, they are inherited by the symmetries of $\Ham_{\even}$ and $\Ham_{\odd}$. Hence, full spin rotational invariance is broken except for $\DeltaY=\DeltaZ=1$; for $\DeltaY=1$, $\Spin_{\tot}^z$ symmetry is preserved and $\Spin_{\tot}^z\ket{\psi_{\Ptrot}}=0$. 
Moreover, since both $\Ham_{\even}$ and $\Ham_{\odd}$ commute with $\Translation^2$, $\Parity_b$ and $\Inversion$, we immediately deduce that: 
\begin{eqnarray*} 
    \Translation^2\ket{\psi_{\Ptrot}} &=& \ket{\psi_{\Ptrot}} 
    \hspace{15mm} 
    \Inversion \ket{\psi_{\Ptrot}} = (-1)^{\frac{N}{2}} \ket{\psi_{\Ptrot}} \nonumber \\
    \Parity_b \ket{\psi_{\Ptrot}} &=& (-1)^{\frac{N}{2}} \ket{\psi_{\Ptrot}}  \;.
\end{eqnarray*}
These are precisely the quantum numbers of the ground states we want to construct for $\Ham_{\XYZ}$, as it can be verified numerically for small-size systems with exact diagonalization.

Restricting the variational wavefunction to the ground state symmetry subsector may foster trainability, but this is not the only practical advantage.
Indeed, also the number of independent Pauli correlators needed to compute the variational energy is reduced.

In order to prove this fact, we need to obtain an explicit formula for the variational energy in Eq.~\eqref{eq:cost_function}. It is useful to introduce the $k$-points correlation functions:
\begin{equation}\label{eq:correlators}
    C^{b_i,b_j,\dots,b_k}_{i, j,\dots,k}\left(\bbeta,\balpha \right)_{\Ptrot} := \bra{\psi\left(\bbeta,\balpha \right)_{\Ptrot}} \PauliSigma_{i}^{b_i} \PauliSigma_{j}^{b_j}\cdots\PauliSigma_{k}^{b_k}\ket{\psi\left(\bbeta,\balpha \right)_{\Ptrot}}\;, 
\end{equation}
where the lower indices enumerate the involved spins $i,j,k=1\cdots N$, while the upper indices assign corresponding directions $b=x,y,z$.
Contrarily to quantum chemistry applications~\cite{VQEsurveyNEW} or some classical optimization problems~\cite{QAOA_perceptron_NEW}, our quantum spin model is $2$-local, so that the expectation value in Eq.~\eqref{eq:cost_function}
only requires calculating two-points correlators of the type
\begin{equation}\label{eq:correlators_XYZ}
    C^{b}_{i, j}\left(\bbeta,\balpha \right)_{\Ptrot} := \bra{\psi\left(\bbeta,\balpha \right)_{\Ptrot}} \PauliSigma_{i}^{b} \PauliSigma_{j}^{b}\ket{\psi\left(\bbeta,\balpha \right)_{\Ptrot}}\;. 
\end{equation}
In addition, $C^b_i=\bra{\psi} \PauliSigma_{i}^{b}\ket{\psi}=0$ thanks to the parity symmetry.
Importantly, we can exploit ansatz symmetries to reduce the number of correlators needed: from the $\Translation^2$ symmetry it immediately follows that 
\begin{eqnarray}
    C^b_{(2j-1), (2j-1)+i}&=C^b_{1, 1+i}\;,\hspace{2.5mm} 
    C^b_{2j, 2j+i}&=C^b_{2, 2+i} \;.
\end{eqnarray}
Additionally, since only nearest-neighbors correlators are needed,
the $\Ham_{\XYZ}$ expectation value reduces to
\begin{equation} \label{eq:var_en_XYZ}
     \frac{2}{N} E_N \left(\bbeta,\balpha\right)_\Ptrot=  \sum_{b=x,y,z}
     C^b_{1, 2}\left(\bbeta,\balpha \right)_{\Ptrot}+C^b_{2, 3}\left(\bbeta,\balpha \right)_{\Ptrot}\;,
\end{equation}
involving only six independent correlators, which may be further reduced to four (two) by exploiting rotational symmetries in the XXZ (XXX) case. 
Moreover, a significant reduction in the number of shots to estimate expectation values in real experiments might be accomplished by a final rotation into the Bell basis at the end of the circuit, allowing to directly access the correlator statistics for $b=x,y,z$ by usual measurements in the computational basis~\cite{gokhale2019minimizing}.

It is relevant to notice that this symmetry-encoding procedure can be applied to a subset of symmetries of $\Ho_{\target}$, but it need not be applied to all of them.
Indeed, for the XYZ models, our ansatz state encodes all the aforementioned symmetries of $\Ho_{\target}=\Ham_{\XYZ}$, whereas it does not encode its one-site translational symmetry.
Nevertheless, the latter is almost-exactly restored for optimal variational parameters, as clearly shown in Appendix~\ref{app:additional_numerical_res} (see Fig~\ref{fig:Inf_XYZ}).

The previous discussion extends straightforwardly to the LTFIM, with a few minor modifications.
This variational ansatz lies in the same symmetry subsector as the target ground state: 
precisely, both are eigenstates with eigenvalue +1 of the symmetry operators $\widehat{\operatorname{T}}$ (full translational invariance) and $\Inversion$.
This fact is once again easily verified, since the initial fully-polarized along-x state $\ket{\psi_0}$ is in the same symmetry sector, and both symmetries commute with the generators of the HVA wavefunction.
Incidentally, note that the LTFIM parameterized quantum circuit reduces to the usual QAOA ansatz for the TFIM ($g_z=0$), where also the parity symmetry $\Parity_x$ is restored and the QAOA ansatz has the same eigenvalue of the ground state (for $g_x>0$).

Along the same lines of the previous discussion, due to the full translational invariance, we now have:
\begin{equation}
\label{eq:correlators_LTFIM}
\begin{split}
C^{z}_{i, j}\left(\bbeta,\balpha \right)_{\Ptrot} &= C^{z}_{1, 2}\left(\bbeta,\balpha \right)_{\Ptrot} \\
C^{b}_{i}\left(\bbeta,\balpha \right)_{\Ptrot} &= C^{b}_{1}\left(\bbeta,\balpha \right)_{\Ptrot}\;,
\end{split}
\end{equation}
for any $b=x,y,z$, leading to the following expression for the variational energy:
\begin{align}
     \frac{1}{N} E_N \left(\bbeta,\balpha \right)_{\Ptrot}= C^{z}_{1, 2}\left(\bbeta,\balpha \right)_{\Ptrot}- \sum_{b=x,z} g{_b}\, C^b_{1}\left(\bbeta,\balpha \right)_{\Ptrot}\;.
\label{eq:var_en_LTFIM}
\end{align}

This reduction in the number of independent correlators --- thanks to symmetry encoding in HVA --- 
is handy for classical simulations but also in realistic experiments on a quantum device~\cite{Kliesch_2021}.


As a final remark concerning additional symmetries, since both the XYZ and LTFIM Hamiltonians are real-valued matrices, and we are using ansatz wavefunctions of the generic form in Eq.~\eqref{eq:QAOA_like_ansatz}, then $E_N  \left(\bbeta,\balpha \right)_{\Ptrot}=E_N  \left(-\bbeta, -\balpha \right)_{\Ptrot}$ (time-reversal symmetry).

\subsection{Alternative HVA implementation}
We remark that the number of variational parameters of our HVA for the XYZ models (defined by Eqs.~\eqref{eq:ansatz_state},~\eqref{eq:ansatz_evolution_XYZ}) is always equal to $2\Ptrot$. This is at variance with a different version of the HVA for the XXZ model, studied e.g. in~\cite{Wiersema_2020}, which introduces more parameters to account for the possible spin anisotropies in the Hamiltonian: here, on the contrary, the $\Ham_{\XYZ}$ spin anisotropies are directly accounted for by $\Ham_{\even}$ and $\Ham_{\odd}$ in Eq.~\eqref{eq:ansatz_evolution_XYZ}, using only two parameters per layer. 
Another possible approach was tested in~\cite{Montenegro}, by adopting a more general class of ansatz wavefunctions, with a number of variational parameters per layer proportional to 
the system size $N$.

Alternative HVA formulations have also been proposed for LTFIM ground state preparation, as those in Ref.~\cite{MatosPRXQuantum.2.010309,LaroccaDiagnBP}.
Once more, here we only need 
two variational parameters per layer, regardless of the phase diagram point in consideration.

\section{The light-cone for XYZ and LTFIM}
\label{app_light-cone}

\begin{figure}[ht]
\centering
\includegraphics[width=0.21\textwidth]{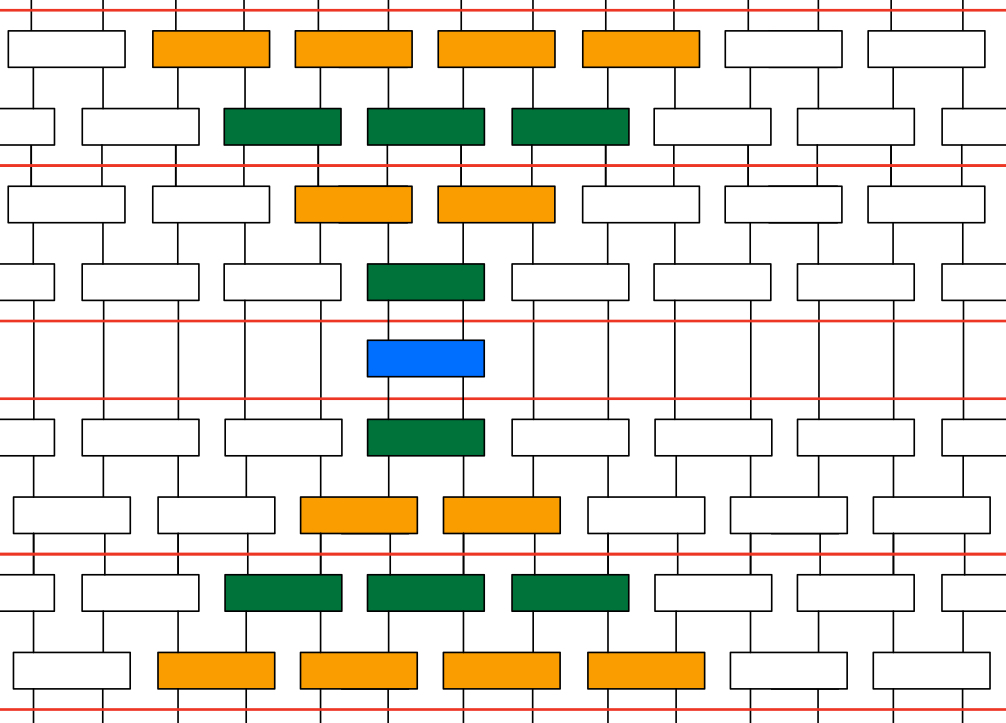}
\includegraphics[width=0.245\textwidth]{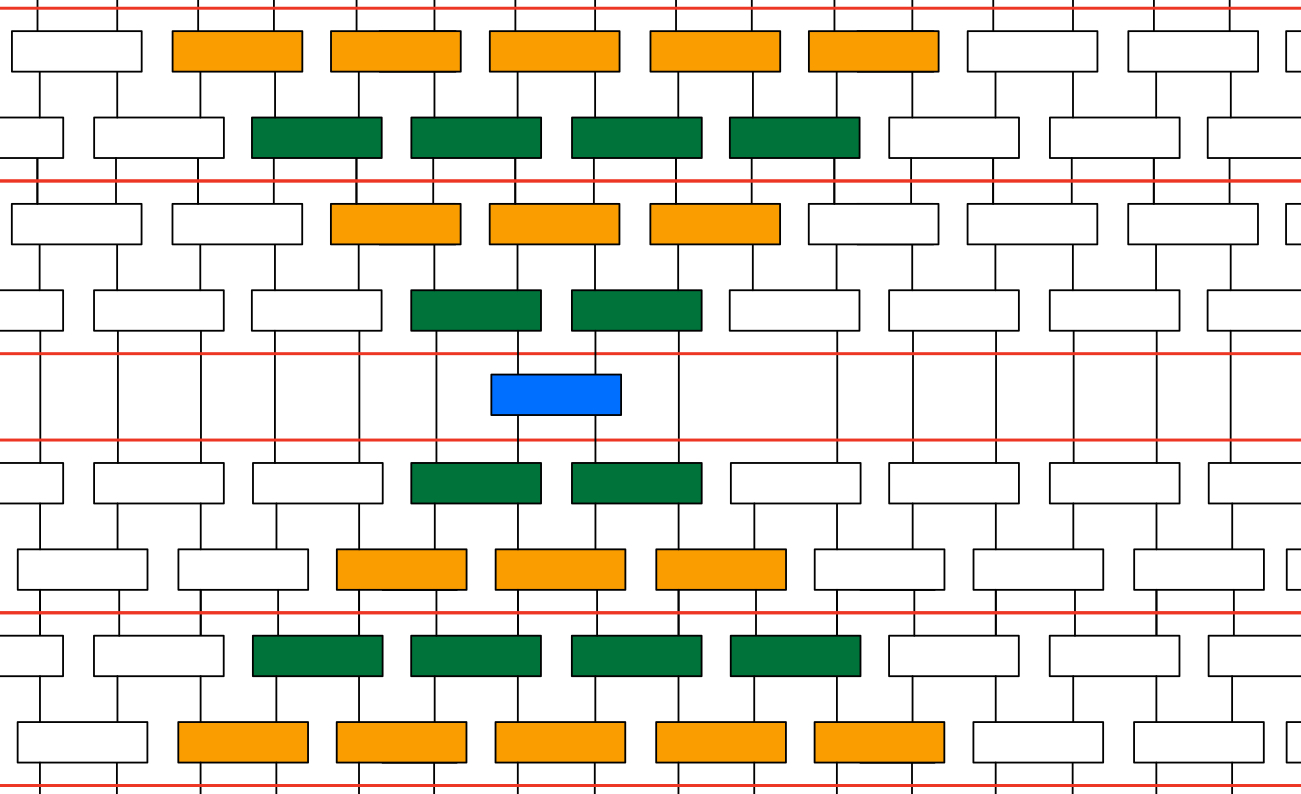}
\caption{The light-cone spreading of quantum correlations for  $C^b_{1, 2}$ (left) and  $C^b_{2, 3}$ (right) for the XYZ ansatz as in Eqs~\ref{eq:ansatz_state},~\ref{eq:ansatz_evolution_XYZ}, with a given depth $\Ptrot$ (here $\Ptrot=2$).
The blue blocks represent the observable, while the orange (green) blocks represent the relevant part of $\nep^{-i\beta_m\Ham_{\even}}$ ($\nep^{-i\alpha_m\Ham_{\odd}}$). Only a reduced spin chain of length proportional to $\Ptrot$ is involved in the calculation: the white blocks (acting on spins outside the reduced chain) can be trivially contracted. The gates are arranged according to a \emph{brickwork}~\cite{ComplexityHaferkamp_2022} architecture.
}\label{fig:light_cone_XYZ}
\end{figure}

In Appendix~\ref{app:models_and_ansatzs} we derived a simplified expression of the variational energy for the XYZ model and the LTFIM, requiring the evaluation of a small set of two-points and one-point correlators, as in Eqs.~\eqref{eq:var_en_XYZ} and~\eqref{eq:var_en_LTFIM}. The calculation of these correlators admits a simple graphical interpretation in terms of a ``light-cone''  spreading of quantum correlations, which is an immediate consequence of the locality of two-body spin interactions, reminiscent of Lieb-Robinson bounds. 

Let us first focus on the XYZ case.
In view of Eq.~\eqref{eq:var_en_XYZ}, it is sufficient to compute only $C^b_{1, 2}$ and $C^b_{2, 3}$: this can be done by addressing a reduced spin chain of length proportional to $\Ptrot$, as sketched in Fig.~\ref{fig:light_cone_XYZ}.
Remarkably, the reduced spin chain is smaller than the whole chain --- and it ``does not see'' the boundary conditions~\cite{Mbeng_arXiv2019} --- only for small-enough values of $\Ptrot$, corresponding to a low-depth quantum circuit for our ansatz.
When this is the case, it can be proven that $C^b_{1, 2}$ and $C^b_{2, 3}$ do not depend on the system size $N$, and neither does the rescaled variational energy in Eq.~\eqref{eq:var_en_XYZ}.
More precisely, within this graphical interpretation, it is easy to observe that $C^b_{1, 2}$ is independent of $N$ if  $4\Ptrot<N$, while for $C^b_{2, 3}$ the condition reads $4\Ptrot+2<N$ (notice the minor differences between the two light-cones in Fig.~\ref{fig:light_cone_XYZ}).
%
Therefore, once we have fixed the depth $\Ptrot$ of the ansatz, for large-enough sizes $N$ of the XYZ chain the whole (rescaled) variational energy landscape defined by Eq.~\eqref{eq:var_en_XYZ} does not depend on $N$. 
Precisely, this holds true for any $N>\tilde{N}_\Ptrot$, where $\tilde{N}_\Ptrot=4\Ptrot+2$.
This analysis implies that the optimal parameters for given $\Ptrot$, found for an XYZ chain of size $N>\tilde{N}_\Ptrot$, can be \emph{exactly} transferred to any chain of size $N'>N$. 

\begin{figure}[ht]
\centering
\includegraphics[width=0.213\textwidth]{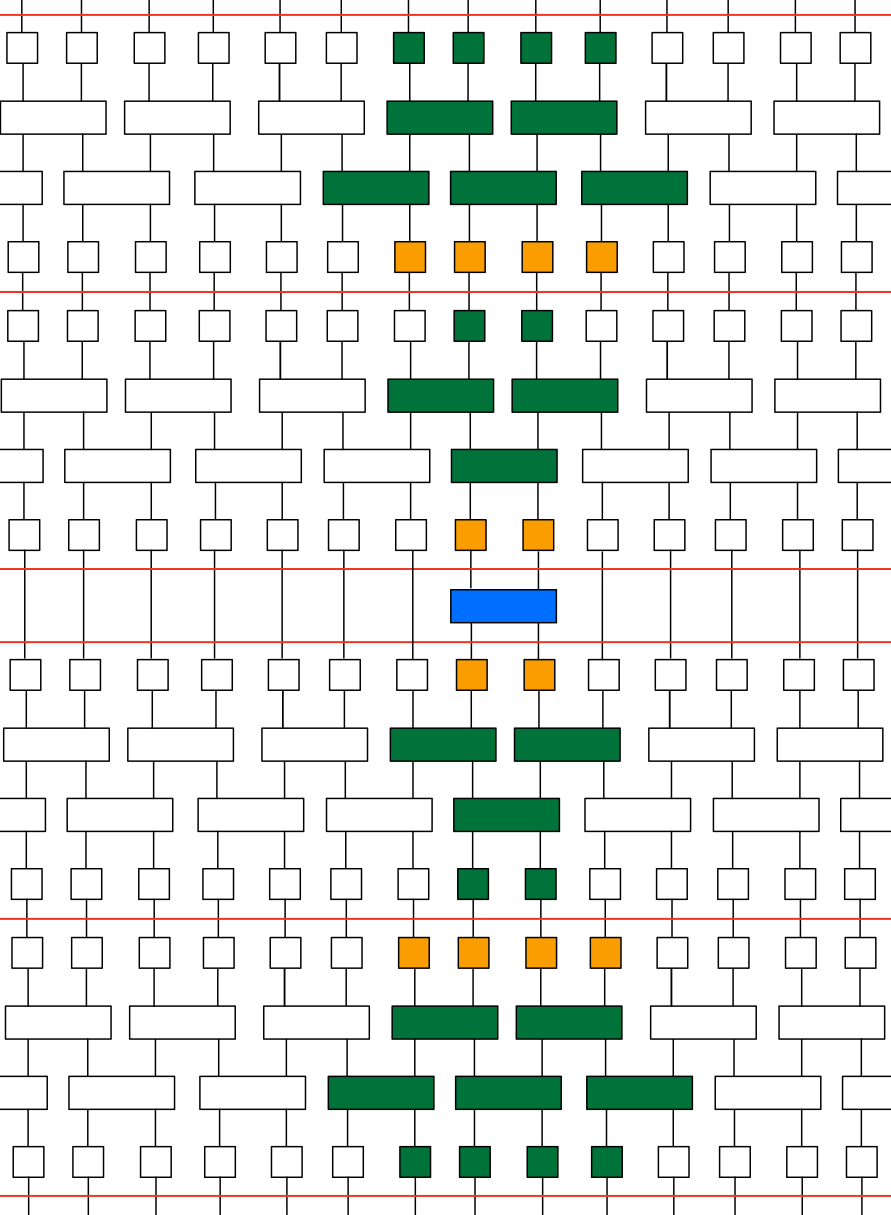}
\includegraphics[width=0.25\textwidth]{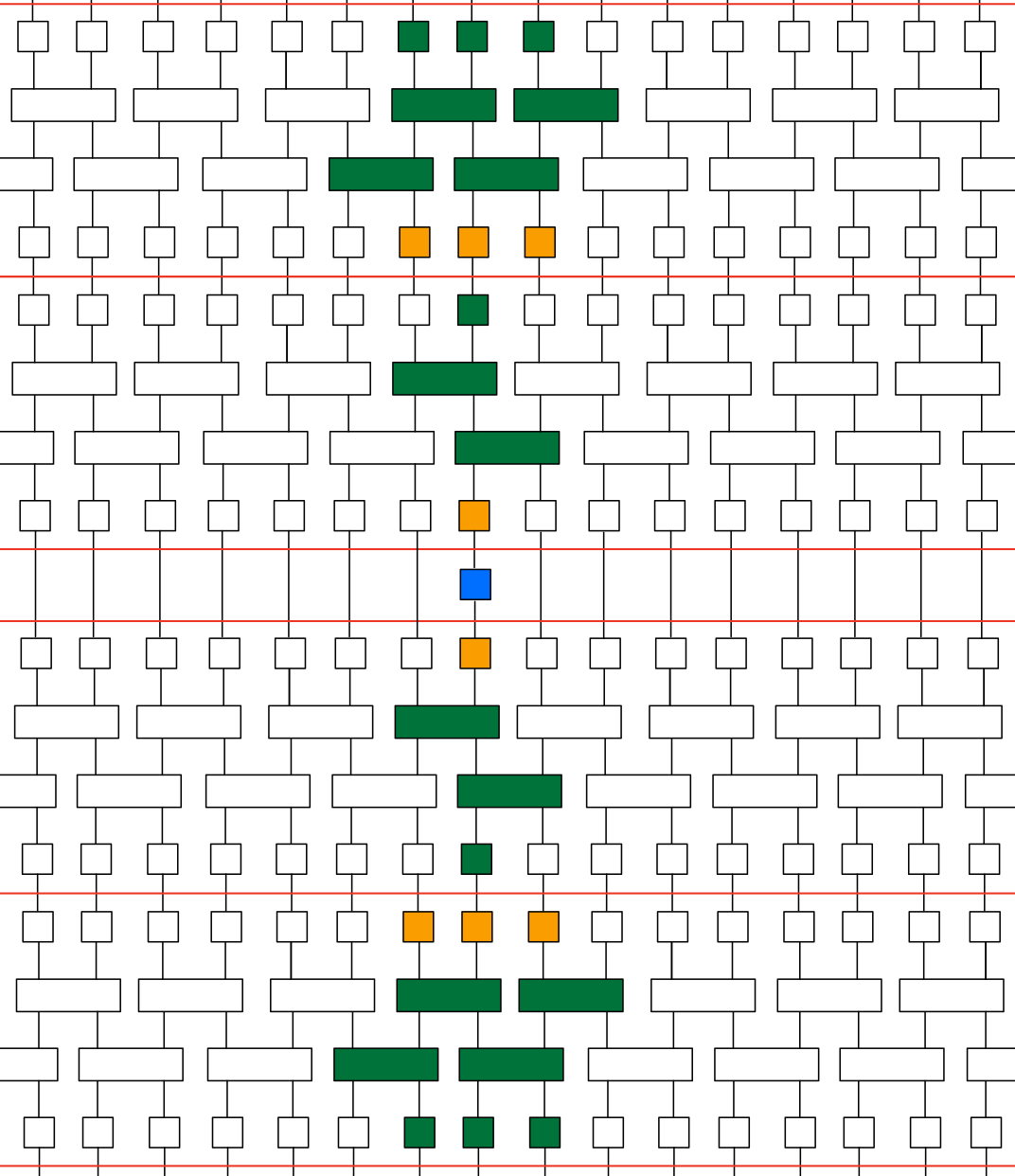}
\caption{The light-cone for  $C^{z,z}_{1, 2}$ (left) and  $C^x_{1}$, $C^z_{1}$ (right) for the LTFIM ansatz as in Eqs.~\eqref{eq:ansatz_state},~\eqref{eq:ansatz_evolution_LTFIM} for a depth $\Ptrot$ (here $\Ptrot=2$). 
The blue blocks represent the observable, while the orange (green) blocks represent the relevant part of $\nep^{i\beta_m \Ham_X }$ ($ \nep^{-i\alpha_m \left(\Ham_{ZZ} - g_z \Ham_Z \right)}$). Also in this case, only a reduced spin chain of length proportional to $\Ptrot$ is involved in the calculation: the white blocks (acting on spins outside the reduced chain) can be trivially contracted. }
\label{fig:light_coneLTFIM}
\end{figure}

Similarly to the XYZ case, a description of quantum correlations spreading in terms of a light-cone emerges also for the LTFIM (analogously to the discussion in~\cite{Mbeng_arXiv2019} for the TFIM).
In particular, one can prove that the rescaled variational energy in Eq.~\eqref{eq:var_en_LTFIM} does not depend on the system size $N$ if $N>\tilde{N}_\Ptrot$, with $\tilde{N}_\Ptrot=2\Ptrot+1$. This follows from the graphical interpretation in Fig.~\ref{fig:light_coneLTFIM}.

\section{Algorithmic details and additional numerical results}
\label{app:additional_numerical_res}

\subsection{Interpolation algorithm (INTERP)}
\begin{figure}
\centering
\includegraphics[width=0.475\textwidth]{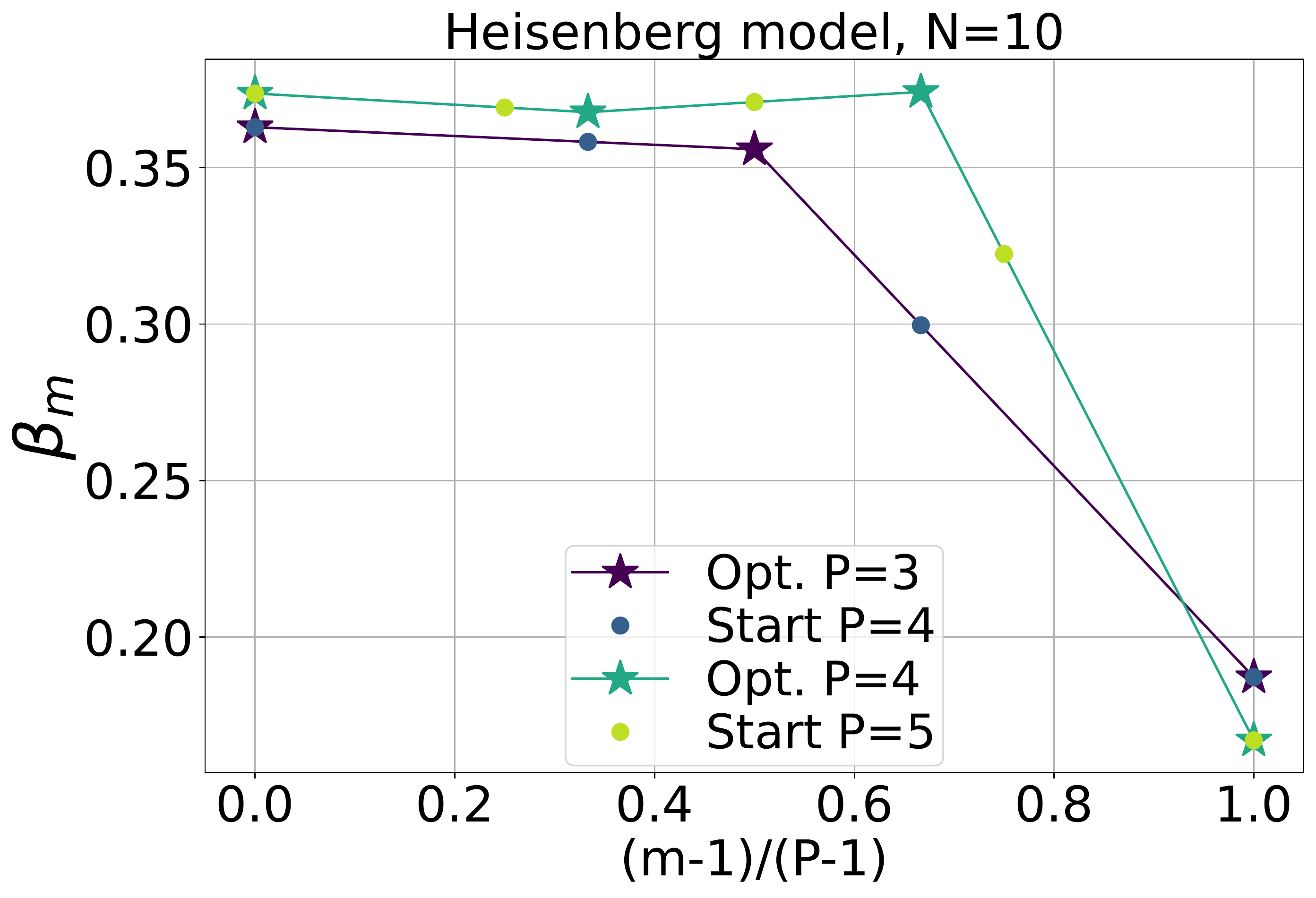}
\caption{Visual example of the interpolation strategy, showing few initial steps of the iterative algorithm that is described in the main text.}
\label{fig:InterpExampleFig}
\end{figure}

Here we describe the INTERP (interpolation) procedure, specifying technical information on our implementation.
The INTERP strategy, introduced in \cite{Zhou_2020}, is an algorithm devised for an iterative optimization of variational parameters of the cost function, originally applied in the context of QAOA for classical combinatorial optimization tasks.
The INTERP strategy works as follows:
\begin{enumerate}
    \item The optimization starts from a guess of the initial parameters at $\Ptrot=\Ptrot_\text{min}$ (e.g. $\Ptrot_\text{min}=1$), namely $\left(\boldsymbol{\beta},\boldsymbol{\alpha}\right)^\text{start}_{\Ptrot=\Ptrot_\text{min}}$.
    \item  We run a local optimization starting from $\left(\boldsymbol{\beta},\boldsymbol{\alpha}\right)^\text{start}_{\Ptrot=\Ptrot_\text{min}}$, using a classical local optimization routine, in order to minimize the cost function and obtain new optimized angles $\left(\boldsymbol{\beta},\boldsymbol{\alpha}\right)^\text{opt}_{\Ptrot=\Ptrot_\text{min}}$.
    \item We run the following instructions on a loop up to $\Ptrot=\Ptrot_\text{max}$:
\begin{enumerate}
    \item Given the optimal parameters at step P, $\left(\boldsymbol{\beta},\boldsymbol{\alpha}\right)^\text{opt}_{\Ptrot}$, we set the initial parameters at step $\Ptrot+1$,  $\left(\boldsymbol{\beta},\boldsymbol{\alpha}\right)^\text{start}_{\Ptrot+1}$, using the interpolation formula of Ref.~\cite{Zhou_2020} for $i=1,2, \ldots, \Ptrot+1$:
\begin{align*}
\qquad \qquad \left[\boldsymbol{\beta}^\text{\,start}_{\Ptrot+1}\right]_{i}=\frac{i-1}{\Ptrot}\left[\boldsymbol{\beta}^\text{\,opt}_{\Ptrot}\right]_{i-1}+\frac{\Ptrot-i+1}{\Ptrot}\left[\boldsymbol{\beta}^\text{\,opt}_{\Ptrot}\right]_{i},
\end{align*}
where $\boldsymbol{\beta}^\text{\,opt}_{\Ptrot}$ is a P-dimension vector.
Note that it is not required to define values of $\left[\boldsymbol{\beta}^\text{\,opt}_{\Ptrot}\right]_{0}$ and $ \left[\boldsymbol{\beta}^\text{\,opt}_{\Ptrot}\right]_{\Ptrot+1}$, since they are multiplied by null coefficients in the formula. The same rule applies to $\boldsymbol{\alpha}$ angles.
\item We run a new local optimization starting from $\left(\boldsymbol{\beta},\boldsymbol{\alpha}\right)^\text{start}_{\Ptrot+1}$, yielding a new set of angles $\left(\boldsymbol{\beta},\boldsymbol{\alpha}\right)^\text{opt}_{\Ptrot+1}$.
\item We increment the value of $\Ptrot$ by one unit: $\Ptrot\rightarrow \Ptrot+1$.
\end{enumerate}
\end{enumerate}
As a visual support, in Fig.~\ref{fig:InterpExampleFig} we sketch the INTERP procedure for a simple case, starting from the optimized parameters $\left(\boldsymbol{\beta},\boldsymbol{\alpha}\right)^\text{opt}_{\Ptrot=3}$, and finding (in sequence) the angles $\left(\boldsymbol{\beta},\boldsymbol{\alpha}\right)^\text{start}_{\Ptrot=4}$, $\left(\boldsymbol{\beta},\boldsymbol{\alpha}\right)^\text{opt}_{\Ptrot=4}$, $\left(\boldsymbol{\beta},\boldsymbol{\alpha}\right)^\text{start}_{\Ptrot=5}$.

In practice, this is the specific version of INTERP we used in this paper, involving a single-unit increase of the value of $\Ptrot$ at each iteration, starting from $\Ptrot=1$. Nevertheless, several minor modifications can be made to this scheme, and whole other iterative methods have also been developed~\cite{VQE_review_NEW}.
Concerning the initial guess at $\Ptrot=1$, we always set $\left(\beta_1,\alpha_1\right)^\text{start}_{\Ptrot=1}$ = $\left(1/10,1/10\right)$, as a starting point to run the first preliminary optimization. Albeit this choice might be arbitrary, in practice we verified that this preliminary optimization always converges to a well-defined minimum in the ($\Ptrot=1$) search space.
Moreover, this minimum is close to the origin, which might provide a useful bias toward short total coherence times in the iterative construction of the smooth optimal curve.

The code for numerical simulations is written with \emph{Qiskit}~\cite{Qiskit_New} (using as classical optimizer the L-BFGS-B algorithm~\cite{L_BFGS_B}).
We test INTERP algorithm by artificially fixing a maximum number of iterations for the classical optimizer: throughout the paper we set $N_\text{iter}=100$, but our results are qualitatively unaffected by moderately reducing (or increasing) this value.
This fixed maximum number of iterations clearly sets an upper bound on the computational resources of the algorithm.
In practice, by stopping the optimization loop, we find quasi-optimal schedules, which, however, are good enough to obtain almost-unit fidelity with the exact ground state.

\subsection{Numerical results on ground state fidelity}

\begin{figure}[ht]
\centering
\includegraphics[width=0.45\textwidth]{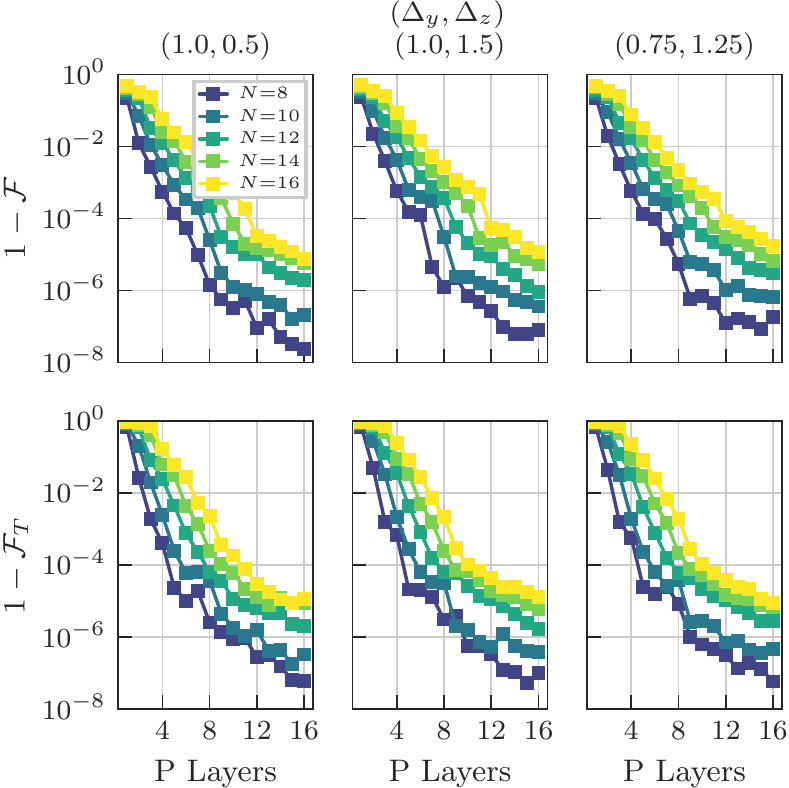}
\caption{Infidelity and translational infidelity for the optimal XYZ ansatz, as a function of increasing number of layers. Even lower (better) values can be obtained for the Heisenberg model ($\DeltaY=\DeltaZ=1$). 
The translational infidelity follows a similar pattern, proving that full translational symmetry is correctly restored.
The optimal parameters always lie on a smooth curve, as shown in Fig.~\ref{fig:smooth_curves}.}
\label{fig:Inf_XYZ}
\end{figure}

\begin{figure}[ht]
\centering
\includegraphics[width=0.45\textwidth]{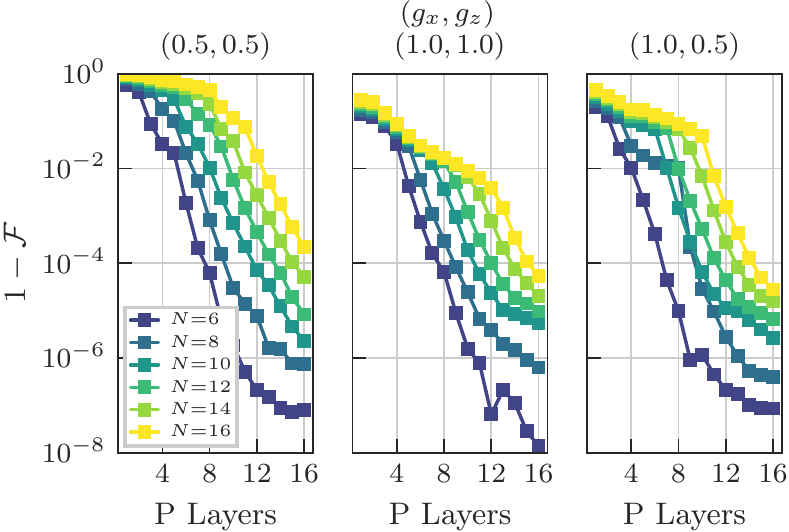}
\caption{Infidelity for the optimal LTFIM ansatz, as a function of increasing number of layers. As for the XYZ models, we  observed empirically that similar infidelities can be obtained for other points of the phase diagram (different values of $g_x$ and $g_z$). Once again, the optimal parameters lie on a smooth curve, as shown in Fig.~\ref{fig:smooth_curves}.}
\label{fig:Inf_LTFIM}
\end{figure}

We tested the effectiveness of INTERP in providing excellent approximate ground states for our models, by applying it to the ansatz wavefunctions defined by Eqs.~\eqref{eq:ansatz_state},~\eqref{eq:ansatz_evolution_XYZ} (Eqs.~\eqref{eq:ansatz_state},~\eqref{eq:ansatz_evolution_LTFIM}) for the XYZ (LTFIM).
We quantify the accuracy of our ground state approximation --- namely the ansatz state evaluated at optimal parameters $(\bbeta^{\qopt},\balpha^{\qopt})_\Ptrot$ --- with the ground state fidelity 
\begin{equation} \label{eq:fidelity}
    \Fidelity_{\Ptrot} = \Big| \langle \psi_{\gs} | \psi(\bbeta^{\qopt},\balpha^{\qopt})_\Ptrot \rangle \Big|^2 \;.
\end{equation}
Another useful quantity for the XYZ model is the {\em translational fidelity} of the optimal state with its one-site translated version:
\begin{equation}
    \Fidelity^{\rmT}_{\Ptrot} = \Big| \langle \psi(\bbeta^{\qopt},\balpha^{\qopt})_\Ptrot | \Translation | \psi(\bbeta^{\qopt},\balpha^{\qopt})_\Ptrot \rangle \Big|^2 \;.
\end{equation}
Indeed, as stated in Appendix~\ref{app:models_and_ansatzs}, the ansatz for the XYZ model does not encode one-site translational symmetry. Nevertheless, this translational fidelity is expected to converge to one when approximating the true ground state with high fidelity, thus restoring the full translational symmetry.

\begin{figure}[ht]
\centering
\includegraphics{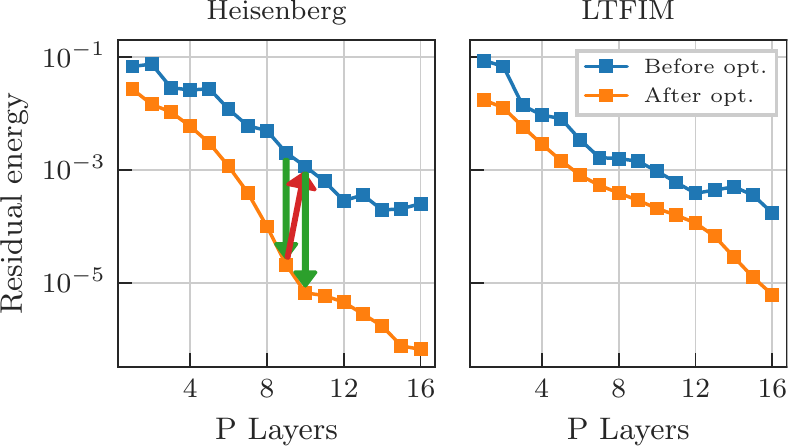}
\caption{Residual energy vs $\Ptrot$ before and after each local optimization leading from $\left(\boldsymbol{\beta},\boldsymbol{\alpha}\right)^\text{start}_{\Ptrot}$ to $\left(\boldsymbol{\beta},\boldsymbol{\alpha}\right)^\text{opt}_{\Ptrot}$. Data refer to $N=16$ qubits, for both the ($\DeltaY=1$, $\DeltaZ=1$) Heisenberg model and the ($g_x=1$, $g_z=1$) LTFIM. Interestingly, the residual energy before each optimization is already quite low, especially for large values of $\Ptrot$. The vertical green arrows show the improvement of the local optimization, whereas the red arrow depicts the interpolation step from $\left(\boldsymbol{\beta},\boldsymbol{\alpha}\right)^\text{opt}_{\Ptrot}$ to $\left(\boldsymbol{\beta},\boldsymbol{\alpha}\right)^\text{start}_{\Ptrot+1}$.}
\label{fig:InterpP}
\end{figure}

Some illustrative results for the ground state fidelity are reported in Fig~\ref{fig:Inf_XYZ} for XYZ models (also displaying data for the translational fidelity) and in Fig~\ref{fig:Inf_LTFIM} for the LTFIM. Note that, for ease of visualization, we plot infidelity (translational infidelity) values, simply defined as
$1-\Fidelity_{\Ptrot}$ ($1-\Fidelity^{\rmT}_{\Ptrot}$).
Remarkably, INTERP method avoids low-quality local minima of the energy landscape~\cite{VQA_traps_NEW}, converging to smooth optimal curves for the variational parameters (see Fig.~\ref{fig:smooth_curves}) bearing high fidelity values. Indeed, this iterative scheme provides an effective
warm start at each iteration of the algorithm, as it can be understood graphically in Fig.~\ref{fig:InterpP}. Here, we plot the residual energy (see definition in Eq.~\eqref{eq:residual_energy}) for increasing values of $\Ptrot$, both evaluated at $\left(\boldsymbol{\beta},\boldsymbol{\alpha}\right)^\text{start}_{\Ptrot}$ (before the local optimization) and at $\left(\boldsymbol{\beta},\boldsymbol{\alpha}\right)^\text{opt}_{\Ptrot}$ (after it).

We remark that the same ansatz (defined by Eqs.~\eqref{eq:ansatz_state},~\eqref{eq:ansatz_evolution_XYZ}) successfully prepares the ground state of XXZ in both phases ($\DeltaZ<1$ and $\DeltaZ>1$, with $\DeltaY=1$), of the Heisenberg model ($\DeltaY=\DeltaZ=1$) and also of XYZ: here we show data for some arbitrary values of $\DeltaY, \DeltaZ$, but we verified that these results extend to different points in the phase diagram.
The same comments apply to LTFIM ansatz (defined by Eqs.~\eqref{eq:ansatz_state},~\eqref{eq:ansatz_evolution_LTFIM}), which is effective in all phases.
These high fidelity values are obtained despite the finite number of iterations ($N_\text{iter}=100$) set for the classical optimizer and could be improved by increasing this value.

\subsection{Additional numerical results on transferability and barren plateaus}

\begin{figure}[ht]
\centering
\includegraphics[width=0.475\textwidth]{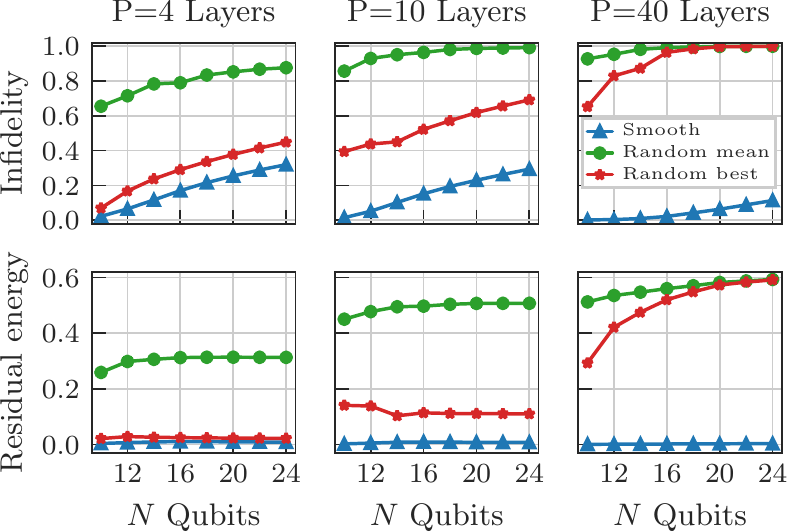}
\caption{Transferability of smooth optimal solutions (INTERP) vs random-start solutions (local optimization), both obtained for a small-size ``guess'' system ($N_G=8$) and then transferred to larger system sizes $N$.
Data refer to the ($\DeltaY=1$, $\DeltaZ=1$) Heisenberg model; we plot the average and the best out of $20$ random-start local optimizations, compared to the same INTERP smooth curve $(\bbeta^*,\balpha^*)\rvert_{\Ptrot,N_G}$ discussed in the main text.
Remarkably, INTERP solutions are always observed to provide an excellent educated guess for larger systems: this is particularly evident for large values of $\Ptrot$, where random-start solutions always fail in this respect.}
\label{fig:transferibilityINF_RANDVSSMOOTH_XXX}
\end{figure}

\begin{figure}[ht]
\centering
\includegraphics[width=0.475\textwidth]{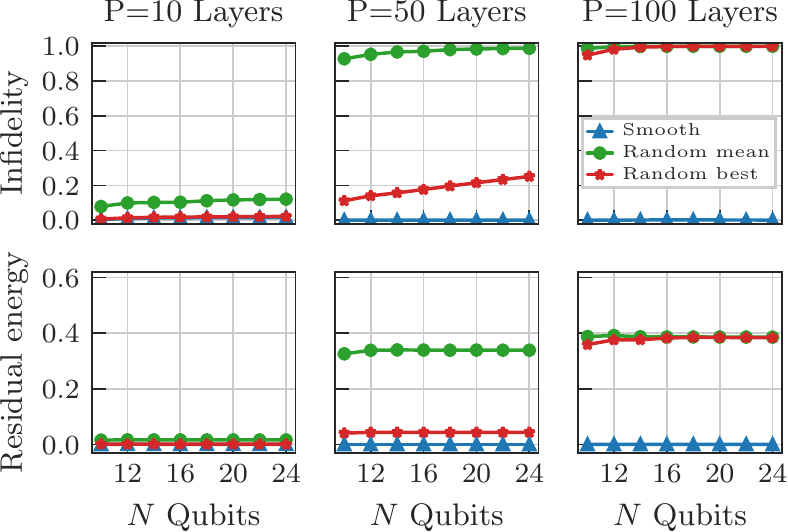}
\caption{Transferability of smooth optimal solutions (INTERP) vs random-start solutions (local optimization), both obtained for a small-size ``guess'' system ($N_G=8$) and then transferred to larger system sizes $N$.
Data refer to the ($g_x=1$, $g_z=1$) LTFIM and we plot the average and the best out of $20$ random-start local optimizations. The same comments apply as in Fig.~\ref{fig:transferibilityINF_RANDVSSMOOTH_XXX}.}
\label{fig:transferibilityINF_RANDVSSMOOTH_LTFIM}
\end{figure}

\begin{figure}[ht]
\centering
\includegraphics[width=0.475\textwidth]{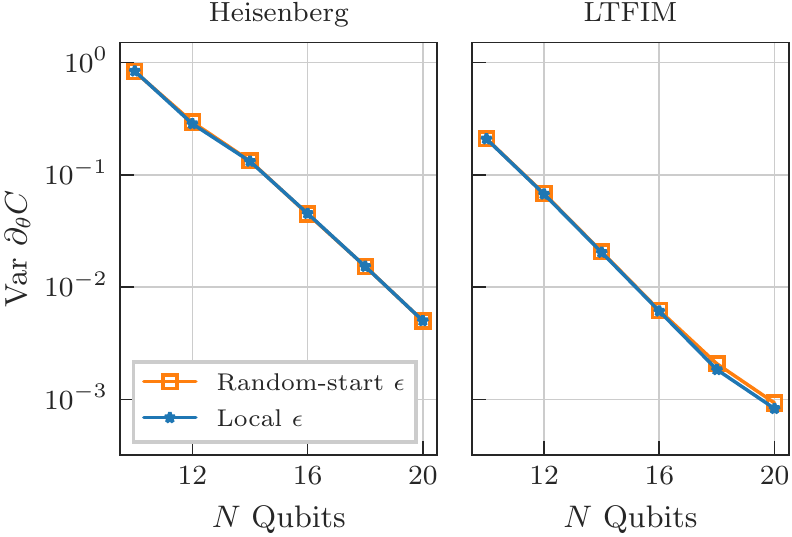}
\caption{Comparison between the local landscape of transferred random-start (non-smooth) solutions with that of random points in the search space (dubbed ``Local $\epsilon$'', same data as in Fig.~\ref{fig:BP2}). In both cases, we average over $20$ instances and, for each instance, we sample $1000$ points in a neighborhood of radius $\epsilon=0.05$. The exponential decay is manifest, with a striking similarity. This is at variance with the local landscape of transferred smooth solutions found via INTERP, not displaying barren plateaus (compare with Fig.~\ref{fig:BP2}).}
\label{fig:BPrandom}
\end{figure}

In the main text, we described the transferability property of a class of smooth solutions found via INTERP.
However, for a small system (``guess size" $N_G=8$) and a large value of $\Ptrot$, one can easily find other solutions by means of standard random-start local optimization, not displaying any smoothness property as a function of the layer index $m=1\cdots\Ptrot$.
Remarkably, these do not offer, in general, any useful educated guess for the ground state preparation of a larger system, as shown in Figs.~\ref{fig:transferibilityINF_RANDVSSMOOTH_XXX} and~\ref{fig:transferibilityINF_RANDVSSMOOTH_LTFIM} for the Heisenberg model and the LTFIM, both in terms of ground state infidelity and residual energy (see Eq.~\eqref{eq:residual_energy}).
We remark that both INTERP and any random-start solution prepare equally well (essentially, with zero infidelity) the ground state for the small system. Nevertheless, only the former class of solutions is observed to always yield an excellent educated guess for the same task with a larger number of qubits. This fact is particularly evident for large values of $\Ptrot$.
Indeed, in this regime, transferred random-start solutions perform as poorly as the ansatz evaluated at random in an arbitrary point of the search space (i.e., almost-unit infidelity, residual energy of the order of $\approx 0.5$).
Moreover, smooth solutions found via INTERP stand out also concerning their favorable local landscape, as discussed in the main text, where gradients do not show any appreciable exponential decay. In contrast, this is not observed in the neighborhood of transferred non-smooth solutions, where barren plateaus are as marked as in the neighborhood of a random point in the search space (or as in the global search space, see Fig.~\ref{fig:BP2}). This result is outlined in Fig~\ref{fig:BPrandom}. Here, and in the rest of the section, we focus again on the ($\DeltaY=1$, $\DeltaZ=1$) Heisenberg model and the ($g_x=1$, $g_z=1$) LTFIM.

\begin{figure}[ht]
\centering
\includegraphics{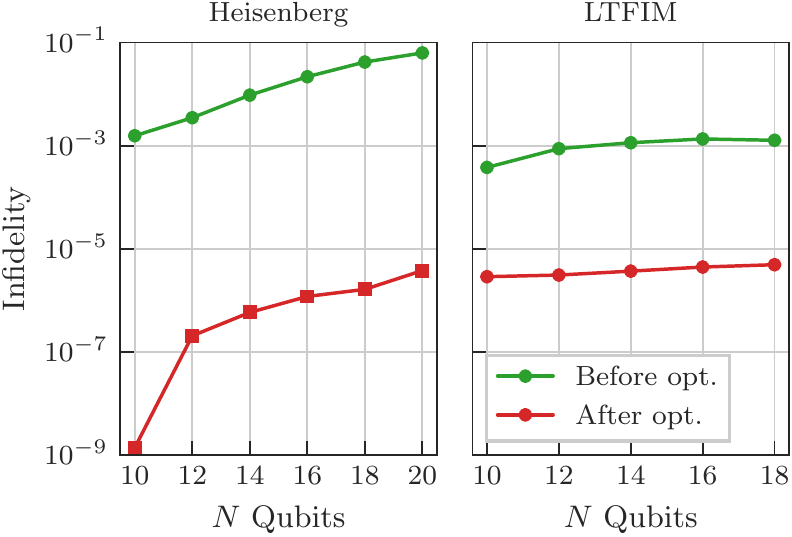}
\caption{Refinement optimization starting from the transferred smooth curve, for the Heisenberg model with $\Ptrot=40$ (left panel) and the LTFIM with $\Ptrot=100$ (right panel). Even with a fixed number of iterations $N_\text{iter}=100$ for the classical optimizer, we can significantly reduce the ground state infidelity.}
\label{fig:RefineOpt}
\end{figure}

\begin{figure}[ht]
\centering
\includegraphics{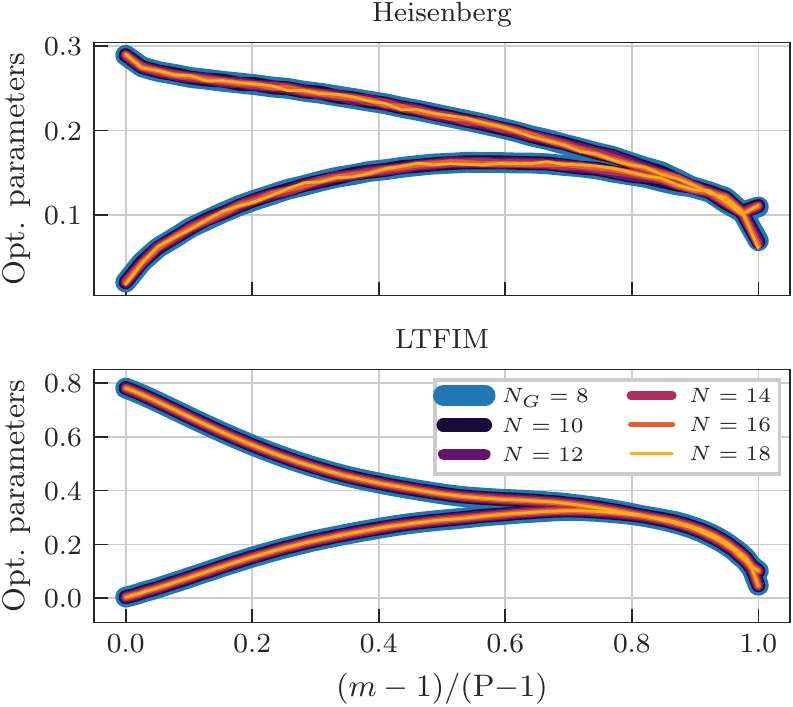}
\caption{A coarse-grained picture of the optimal parameters after performing a refinement optimization, compared with the initial smooth guess (obtained by INTERP) at $N_G=8$, i.e. $(\bbeta^*,\balpha^*)\rvert_{\Ptrot,N_G}$ (see main text). These are the refined optimal parameters corresponding to Fig.~\ref{fig:RefineOpt}, i.e. $\Ptrot=40$ ($\Ptrot=100$) for the Heisenberg model (LTFIM).}
\label{fig:RefineOptSmooth}
\end{figure}

Thanks to this favorable local landscape, one can effectively perform a refinement optimization for the large system, so as to further increase the ground state fidelity above a target threshold. An example is shown in Fig.~\ref{fig:RefineOpt}, by setting a maximum number of iterations $N_\text{iter}=100$ for the classical routine performing the refinement optimization. Despite this constraint, the local optimization succeeds, significantly lowering the final infidelity.
Incidentally, we verified that the refined optimal curve is still in the ``basin of attraction" of the transferred smooth curve, as shown in  Fig.~\ref{fig:RefineOptSmooth}. At this level of detail, the curves appear almost exactly overlapping.

\section{Large-scale simulations for the TFIM}
\label{app:TFIM}

In this section we address the line $(g_z=0)$ of the LTFIM phase diagram, corresponding to the integrable Transverse Field Ising Model (TFIM), with Hamiltonian
\begin{equation} \label{eq:TFIM_Ham}
    \widehat{H}_{\scriptscriptstyle\mathrm{TFIM}}=
    \sum_{i=1}^{N}\hat{\sigma}_{i}^{z} \hat{\sigma}_{i+1}^{z} - g_x\sum_{i=1}^{N}\hat{\sigma}_{i}^{x} \;.
\end{equation}
A mapping to non-interacting fermions allows us to perform large-$N$ VQA simulations, well beyond the usual limits of exact diagonalization techniques. 
In particular, we used the same ansatz as in Eqs.~\eqref{eq:ansatz_state},~\eqref{eq:ansatz_evolution_LTFIM}, which reduces to the standard QAOA ansatz for $g_z = 0$.

As argued in~\cite{LaroccaDiagnBP}, we verified numerically that the TFIM ground state preparation is not affected by barren plateaus.
However, the possibility of large-scale simulations offers a useful benchmark on the effectiveness of INTERP in this regime, in particular concerning the existence of smooth optimal solutions and their transferability.

The variational energy is given by Eq.~\eqref{eq:cost_function} with $\Ho_{\target}=\widehat{H}_{\scriptscriptstyle\mathrm{TFIM}}$, while the residual energy reads as in Eq.~\eqref{eq:residual_energy}, now evaluated in a generic point $(\bbeta,\balpha)_\Ptrot$ of the search space.

Refs.~\cite{Mbeng_arXiv2019,WangFermQAOA,AvoidingXXZtfimWierichs_2020} previously discussed how to efficiently simulate QAOA using the fermionic mapping of the TFIM. They showed that, after applying a Jordan-Wigner and a Bogoliubov  transformation, a system of an even number of spins $N$ decomposes in a direct sum of $N/2$ independent two level systems, which are labelled by the wave-vectors $k_n = \pi \frac{2n-1}{N}$ with $n=1\cdots N/2$. The total residual energy of the one dimensional TFIM then reads
\begin{equation}  \label{eqn:eres_pbc_app}
\varepsilon_N(\bbeta, \balpha)_{\Ptrot} = \sum_{n=1}^{N/2} \varepsilon^{(k_n)}(\bbeta, \balpha)_\Ptrot\;,
\end{equation}
where  $\varepsilon^{(k_n)}(\bbeta, \balpha)_{\Ptrot}$ are the residual energies associated to each individual two-level system. The analytical expression for $\epsilon^{(k_n)}(\bbeta, \balpha)_{\Ptrot}$, as provided in  Ref.~\cite{Mbeng_arXiv2019}, is
\begin{eqnarray}
\epsilon^{(k)}(\bbeta, \balpha)_{\Ptrot} = \frac{1}{2} -\frac{1}{2}\boldsymbol{v}^T_{k} \left( \Tprod{\Ptrot}_{m=1}  R_{\hat{\boldsymbol{z}}}(4 \beta_m )R_{\boldsymbol{b}_k}(4 \alpha_m) \right) \hat{\boldsymbol{z}}\;, \nonumber\label{eqn:pseudospin_full_rotation}
\end{eqnarray}
where 
$\hat{\boldsymbol{z}}=(0,0,1)^T$, $\boldsymbol{b}_k=(-\sin k,0, \cos k)^T$ and   $\boldsymbol{v}_{k} = (\boldsymbol{b}_k + g_x\hat{\boldsymbol{z}})/||\boldsymbol{b}_k + g_x\hat{\boldsymbol{z}}||$ are three-dimensional unit vectors. $R_{\hat{\boldsymbol{\omega}}}(\theta)$ is the $3\times3$ matrix associated with a rotation of 
an angle $\theta$ around the unit vector
$\hat{\boldsymbol{\omega}}$, and their product is ``time''-ordered from right to left for increasing $m=1\cdots\Ptrot$.

\begin{figure}[ht]
\centering
\includegraphics[width=0.47\textwidth]{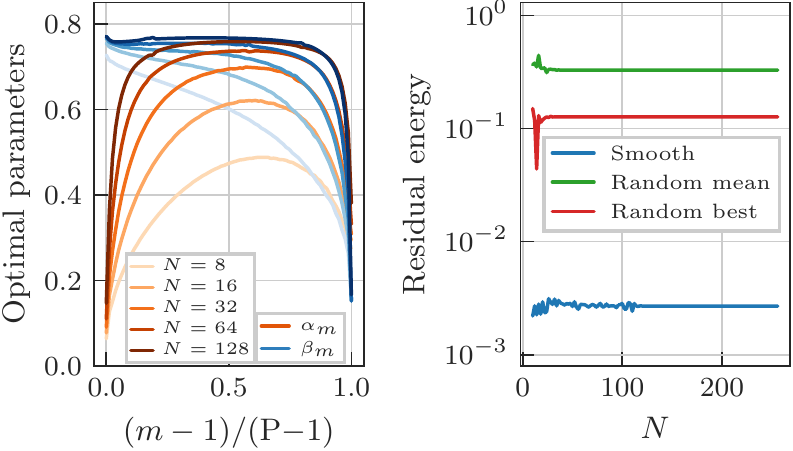}
\caption{(Left panel) Smooth optimal parameter curves for the TFIM, obtained by applying INTERP up to $\Ptrot=80$.
Smooth curves are qualitatively similar for a wide range of system sizes.
(Right panel) Numerical evidence on the transferability of smooth optimal solutions --- obtained for a small-size ``guess'' system ($N_G=8$) --- up to very large systems. In contrast, non-smooth solutions found via random-start local optimization do not provide, on average, any useful guess for the ground state preparation of a larger system, yielding a residual energy $\approx 0.5$.
We averaged over $100$ random-start solutions and, remarkably, not even the best of them is nearly comparable with the one obtained via INTERP. Data refer to the critical point $g_x=1$, and the same behavior is observed in other regions of the TFIM phase diagram.}
\label{fig:smooth_curvesTFIM}
\end{figure} 

These formulas allow for efficient computation of $\varepsilon(\bbeta, \balpha)_{\Ptrot}$, enabling us to numerically study the performance of QAOA for a large number of qubits.
Also for the TFIM we find smooth curves, which are shown in Fig.~\ref{fig:smooth_curvesTFIM} at the critical point $g_x=1$, up to sizes as large as $N=128$.
In the same figure, we also show that smooth curves --- prepared by applying INTERP to a small system of size $N_G=8$ --- are transferable up to sizes as large as $250$ qubits, i.e. they offer a good educated guess for TFIM ground state preparation.
This is in stark contrast with other (non-smooth) solutions, found via random-start local optimization for the small system: despite preparing the small-size ground state with perfect accuracy (the same as applying INTERP), they do not provide any useful educated guess for the large system.

\end{document}